\documentclass[preprint,12pt, 3p]{elsarticle}

\usepackage{amssymb}
\usepackage{caption}
\usepackage[dvipsnames]{xcolor}

\biboptions{numbers,sort&compress}

\newcounter{bla}

\journal{Computer Physics Communications}

\begin{document}

\begin{frontmatter}



\title{Genarris 2.0: A Random Structure Generator for Molecular Crystals}


\author[a]{Rithwik Tom}
\author[b]{Timothy Rose}
\author[b]{Imanuel Bier}
\author[b]{Harriet O'Brien}
\author[c]{́Alvaro V́azquez-Mayagoitia}
\author[a,b,d]{Noa Marom \corref{cauthor}}

\cortext[cauthor] {Corresponding author.\\\textit{E-mail address:} nmarom@andrew.cmu.edu}
\address[a]{Department of Physics, Carnegie Mellon University, Pittsburgh, PA, 15213, USA}
\address[b]{Department of Materials Science and Engineering, Carnegie Mellon University, Pittsburgh, PA, 15213, USA}
\address[c]{Argonne Leadership Computing Facility, Argonne National Lab, Lemont, Illinois 60439, USA}
\address[d]{Department of Chemistry, Carnegie Mellon University, Pittsburgh, PA, 15213, USA}

\begin{abstract}





Genarris is an open source Python package for generating random molecular crystal structures with physical constraints for seeding crystal structure prediction algorithms and training machine learning models. Here we present a new version of the code, containing several major improvements. A MPI-based parallelization scheme has been implemented, which facilitates the seamless sequential execution of user-defined workflows. A new method for estimating the unit cell volume based on the single molecule structure has been developed using a machine-learned model trained on experimental structures. A new algorithm has been implemented for generating crystal structures with molecules occupying special Wyckoff positions. A new hierarchical structure check procedure has been developed to detect unphysical close contacts efficiently and accurately. New intermolecular distance settings have been implemented for strong hydrogen bonds. To demonstrate these new features, we study two specific cases: benzene and glycine. For all polymorphs, the final pool either contained the experimental structure, or structures with similar lattice parameters, symmetry, and packing motifs.

\end{abstract}


\end{frontmatter}


\newpage

{\bf NEW VERSION SUMMARY}

\begin{small}
\noindent
{\em Manuscript Title:} Genarris 2.0: A random structure generator for molecular crystals\\
{\em Authors:} Rithwik Tom, Timothy Rose, Imanuel Bier, Harriet O'Brien, Alvaro Vazquez-Mayagoitia, Noa Marom\\
{\em Program Title:} Genarris 2.0                                          \\
{\em Journal Reference:}                                      \\
{\em Catalogue identifier:}                                   \\
{\em Licensing provisions:} BSD-3 Clause                    \\
{\em Programming language:} Python, C                                   \\
{\em Operating system:} Linux                                 \\
{\em Classification:} Crystallography                                  \\
{\em External routines/libraries:} Spglib, ASE, pymatgen, SciPy, mpi4py, scikit-learn, PyTorch, FHI-aims.                    \\
{\em Nature of problem:} Molecular crystal structure prediction. \\
   \\
{\em Solution method:} Genarris 2.0 generates molecular crystal structures over the 230 space groups, on general and special Wyckoff positions, using physical constraints. Subsampling of the generated structures based on a molecular crystal packing descriptors and an unsupervised machine learning algorithm can be followed by ab initio structure relaxation providing a full-fledged crystal structure prediction workflow. Genarris may also be adapted to the generation of diverse molecular crystal datasets for evolutionary algorithms or machine learning. \\
   \\
{\em Restrictions:} For crystal structure generation, the molecule of interest must be semi-rigid with no bond rotational degrees of freedom. \\
   \\
{\em Unusual features:} Genarris 2.0 is a highly distributed program, making use of MPI for Python to implement bindings of the Message Passing Interface (MPI) and offers the user the ability to design and implement workflows by executing a user-defined list of procedures. Genarris 2.0 implements new features including a machine learning model for estimating the molecular volume in the solid state from the single molecule structure, structure generation in special Wyckoff positions of space groups, hierarchical structure checks including rigorous treatment of non-orthogonal structures, and clustering and down-selection workflows combining first principles simulations with machine learning. \\


\end{small}

\newpage

\section{Introduction}
\label{section:Introduction}

\par The properties of molecular crystals depend not only on their constituents but also the relative arrangement of the molecules inside the unit cell. Properties such as the stability \cite{griesser2008conformational,nyman2015static,hoja2018first}, electronic conductivity \cite{bredas2007charge,jurchescu2009effects,matsukawa2010polymorphs,diao2014understanding,yang2018large}, solubility and bioavailability \cite{ritonavir,disappearing}, have all been observed to vary as a function of the molecular crystal solid state form. The molecules comprising these crystals are held together by weak intermolecular interactions \cite{reilly2015van,hermann2017first} and thus can commonly be experimentally synthesized in multiple forms \cite{stahly2007diversity,lee2011crystal}. This phenomena, known as polymorphism, has been of great importance to pharmaceutical research and for the design of high performance organic electronics \cite{jurchescu2009effects,schrier2011,diao2016polymorphism}. 

\par The field of crystal structure prediction (CSP) is devoted to the prediction of the solid state forms of a molecule \cite{blind1_2000, blind2_2002, blind3_2005, blind4_2007, blind5_2011, blind6_2016, price2014predicting}. CSP requires algorithms that can efficiently generate new structures in order to sample the high dimensional configuration space associated with molecular crystals \cite{day2011current,price2014predicting}. Random, and quasi-random, sampling of the configuration space has been established as a critical component of CSP workflows within the Cambridge Crystallographic Data Centre (CCDC) CSP blind test \cite{blind1_2000, blind2_2002, blind3_2005, blind4_2007, blind5_2011, blind6_2016}. Most of the groups that participated in the sixth CSP blind test used a random crystal structure generation method \cite{DayCSP, PickardCSP, karamertzanis2005, karamertzanis2007, Eijck2000structure}. Random crystal structure generation methods identified four of the five, chemically diverse target systems in the sixth blind test, demonstrating their importance for CSP \cite{blind6_2016}. 

\par Random crystal structure generation methods for CSP follow a similar procedure. First a space group \cite{wonderatschek2004international} is chosen for the new structure. Second, random unit cell parameters commensurate with the space group's crystal system are generated. Third, the molecule positions and orientation of each independent molecule are randomly sampled within the asymmetric unit. Finally, the symmetry operations of the space group are applied to the asymmetric unit generating all molecules in the unit cell. The generated structures are subsequently relaxed using either a system specific force field \cite{mooij1999transferable, price2010modelling, zhang2018harnessing} or a fully \textit{ab initio} approach \cite{li2018genarris, zilka2017ab, PickardCSP}. The success of random structure generation stems comes from unbiased and diverse sampling covering the potential energy surface, followed by a structural relaxation to the nearest local minima, hopefully converging to all experimentally observed polymorphs \cite{PickardCSP,karamertzanis2005,DayCSP}.

\par Despite their overall similarity,, structure generation methods from the sixth blind test differ in subtle ways. Structure parameters may be sampled using either a uniformly random number generator \cite{li2018genarris, PickardCSP}, or quasi-random, low discrepancy sequences \cite{DayCSP,karamertzanis2005,karamertzanis2007}. Structure generation may be performed over all space groups \cite{li2018genarris}, or using only the most common space groups \cite{karamertzanis2005, day2011current} observed in the Cambridge Structural Database (CSD) \cite{allen2002cambridge, groom2016cambridge}. A critical component of the generation procedure is approximating the volume of the molecular crystal before generation. Several methods have been proposed, such as adding up atomic volumes \cite{PickardCSP}, using the morphology of the molecule \cite{DayCSP,pidcock2004novel}, or relaxing a few handmade structures \cite{karamertzanis2005,Eijck2000structure}. It has been demonstrated that random CSP methods may be sensitive to the choice of unit cell volume \cite{DayCSP}. Therefore, it's important to use an accurate volume estimation method. Additionally, structures with reasonable densities are typically closer to their respective local minima making structure relaxations more efficient. Lastly, most random crystal structure generation packages are only capable of generating structures in general Wyckoff positions and rely on the serendipitous generation of structures with molecules occupying special positions.  However, analysis of the CSD has shown that molecules with internal symmetry often occupy special Wyckoff positions \cite{pidcock2003database}. 
 
\par Here we present a new version of Genarris \cite{li2018genarris}, an open source Python package that performs random structure generation for homomolecular molecular crystals of semi-rigid molecules with no bond rotational degrees of freedom using general and special Wyckoff positions. Genarris 2.0 offers several improvements over the previous version. The parallelization model has been changed from Python multiprocessing to MPI for Python (mpi4py) \cite{dalcin2008mpi} to enable more efficient utilization of many cores and seamless sequential execution of user-defined workflows. A new machine learning method for volume estimation, based on a topological molecular descriptor, provides accurate volume predictions across a chemically diverse dataset from the CSD. The speed of structure generation has been significantly increased by developing a new hierarchical scheme for intermolecular distance checks. New settings have been implemented to improve structure generation for systems with strong hydrogen bonds. The performance of new the features in Genarris 2.0 is demonstrated for glycine, which contains relatively strong intermolecular hydrogen bonds, and benzene, a symmetric molecule occupying special Wykckoff position with an inversion center.


\section{Code description}
\label{section:Code Description}

Genarris 2.0 is written in Python 3, with the exception of the new structure generation function, Pygenarris, which is written in C and automatically compiled and installed into Genarris 2.0 as a Python library. Genarris only requires standard Python libraries to install on any machine (i.e. numpy, sci-kit learn, mpi4py, spglib, pymatgen,and ASE, PyTorch). Genarris 2.0 is parallelized with MPI, using mpi4py (Sec.~\ref{subsection:Parallelization}). Pygenarris has additional built in OpenMP support. For energy evaluations and geometry relaxations, Genarris currently interfaces with the electronic structure package FHI-aims \cite{blum2009ab}. It may be adapted to interface with any other electronic structure, force field, or machine learning package that accepts an MPI communicator as an argument.

\begin{figure}[h]
    \centering
    \includegraphics[width=0.35\textwidth]{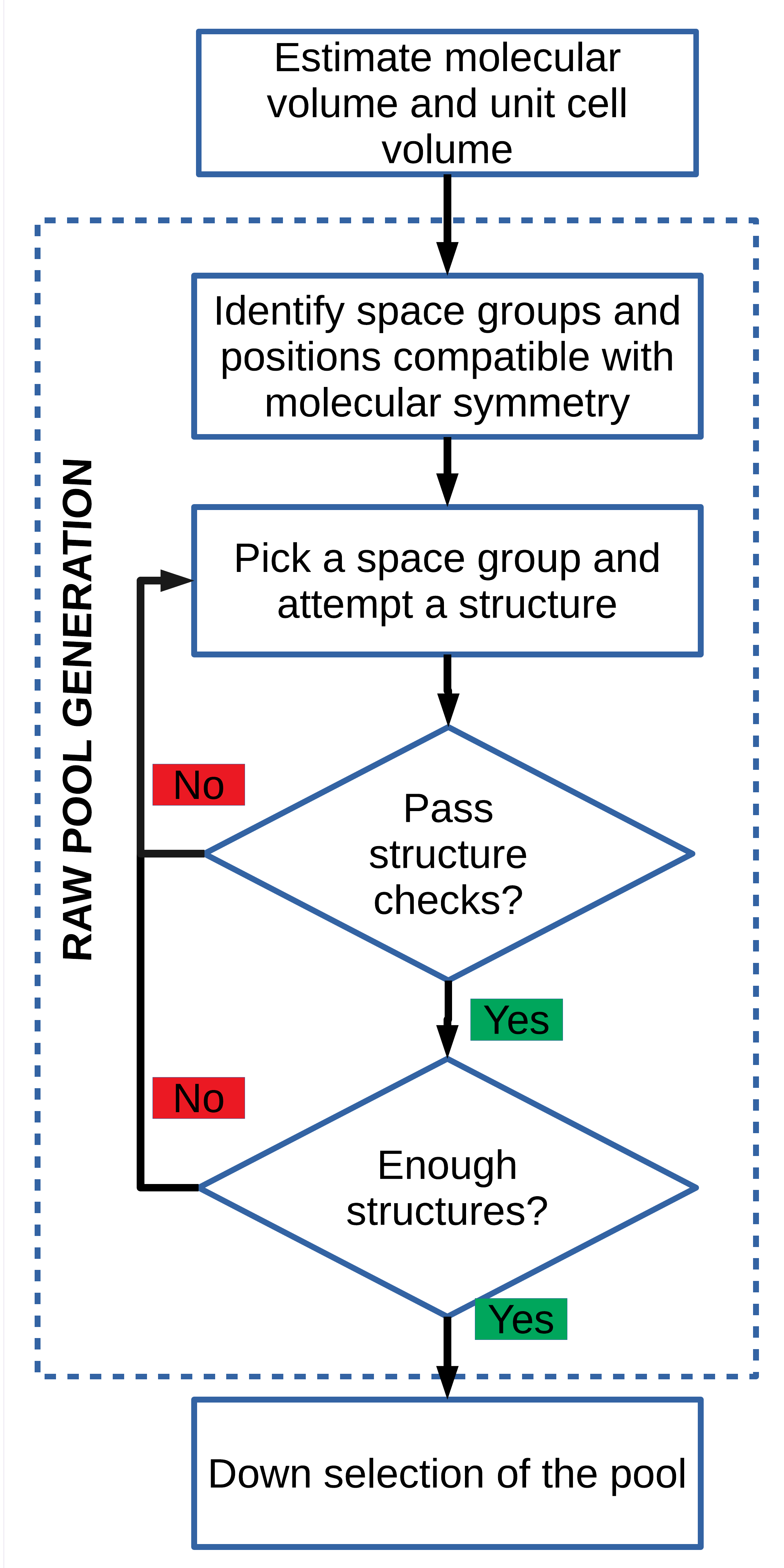}
    \caption{The workflow of Genarris 2.0}
    \label{fig:raw_pool}
\end{figure}

The workflow of Genarris 2.0 is depicted in Figure \ref{fig:raw_pool}.
It begins by estimating the crystal unit cell volume. Given the desired number of molecules per unit cell (Z), the estimate is obtained by relaxing the single molecule geometry and applying a machine-learned model trained on a dataset of experimental structures from the Cambridge Structural Database (CSD) (Sec.~\ref{subsection:Unit cell volume estimation}). Crystal structure generation begins by determining all compatible space groups given the molecular geometry within a user-defined symmetrization tolerance (Sec.~\ref{subsection:Structure generation}). Genarris moves sequentially through this list of space groups, generating a user-defined number of structures per allowed space group and checking them to ensure that no two molecules are unphysically close to each other (Sec.~\ref{subsubsection:Structure checks}). If the user-defined maximum number of consecutive failed generation attempts for a space group is reached, Genarris will proceed to the next space group on the list.

Once a ``raw" pool of physically reasonable, random structures is generated, a user-defined sequence of energy evaluation, clustering, and selection steps may be performed to produce a smaller curated pool of structures, which can be used, e.g., as an initial population for a genetic algorithm \cite{curtis2018gator,curtis2018evolutionary}. For clustering, Genarris uses the affinity propagation (AP) machine learning algorithm \cite{frey2007clustering}. Two types of feature vectors are available in Genarris 2.0, the relative coordinate descriptor (RCD) \cite{li2018genarris} and a radial distribution function (RDF), implemented in PyTorch, similar to that described by Ref. \cite{RDF_1}. Three workflows for down selection have been proposed previously \cite{li2018genarris}. Here, a new ``Robust" workflow is proposed (Sec.~\ref{subsection:Clustering and down-selection}). Lastly, full geometry relaxation may be performed for the final pool of structures.

Genarris 2.0 automatically executes all the procedures in the user-defined procedure list in the order specified. A single input file contains the user defined settings for all desired procedures. This includes the number of cores to be used for each procedure, as different procedures scale differently (see Sec.~\ref{subsection:Parallelization}). Genarris can infer some parameters from previous sections of the workflow. For example, the output file containing the relaxed geometry of the single molecule becomes the default molecule path of subsequent sections if it exists. The user may reorder the procedures as long as the dependencies are satisfied (e.g., feature vector calculation must be performed prior to clustering).

If Genarris is restarted, it determines which step in the procedure list was not completed and resumes from that point. Restarts for generation are implemented by parsing the geometry output file for the last generated space group, and continuing from there. Finished RCD calculations are output as files which permits easy restarts. Restarts for RDF calculations are unnecessary because they take less than a minute for several thousand structures. Genarris supports restarts of FHI-aims jobs by first determining which of them are incomplete and then automatically relaunching them from the last relaxation step.

\subsection{Parallelization}
\label{subsection:Parallelization}
Genarris 2.0 is parallelized using the message passing interface (MPI) paradigm via the mpi4py package. MPI enables immediate cross-platform portability without code changes. The structure generation function in Genarris 2.0 determines the number of allowed space groups for the given molecule, $n$, and accepts as input the number of structures to generate for each of these space groups. Hence, structure generation and subsequent structure checks (Sec.~\ref{subsection:Structure generation}) are embarrassingly parallelized over the total number of structures desired, $N$, with the problem size (maximum number of usable cores) for the generation and structure check procedures equal to $N/n$.

For clustering (see Sec.~\ref{subsection:Clustering and down-selection}), both the RCD and RDF feature vector calculations are embarrassingly parallelized with problem size $N$. The number of clusters produced by the affinity propagation (AP) algorithm \cite{frey2007clustering} is nearly-directly correlated with its \textit{preference} hyperparameter value, but its functional form is not known \textit{a priori}. Therefore, a parallelized version of the standard binary search algorithm has been implemented to output a specified number of clusters $C$ within a tolerance $tol$. The \textit{preference} range is initially wide ($[-1000,1000]$). This range is evenly partitioned into $R$ \textit{preference} values, where $R$ is the number of total MPI ranks available. Each rank executes AP with its assigned value of \textit{preference} and reports the number of clusters obtained to the root rank. The root rank sets the \textit{preference} range upper (lower) bound to the \textit{preference} that returned the lowest (highest) number of clusters above (below) the target number of clusters. The root then partitions the updated \textit{preference} range and assigns each rank its new \textit{preference} value. The procedure is repeated until a \textit{preference} value is found, which yields $C\pm tol$ clusters. The communication required in each iteration is less expensive than an AP call by orders of magnitude, therefore it does not significantly affect the scaling. Because \textit{preference} and the number of clusters are only \textit{nearly}-directly correlated, fail-safes have been implemented. For example, if the current \textit{preference} range fails to yield a number of clusters within $C\pm tol$, then the \textit{preference} range is widened by a random amount. In addition, the user may have the program output the closest number of clusters to $C$ within a desired number of iterations. The memory usage is kept manageable by writing and accessing the affinity and distance matrices via memory maps so that each rank does not make a redundant copy.

Genarris currently interfaces with FHI-aims for energy evaluations and geometry relaxations. It may be adapted to interface with any other electronic structure, force field, or machine learning package that accepts an MPI communicator as an argument. FHI-aims is compiled as a shared library and made an importable Python library through the standard f2py function. The Python communicator is then converted into a Fortran communicator via the py2f method. Thus, Genarris 2.0 allows all ranks to be utilized by FHI-aims. In addition, the world communicator may be split into a user-defined number of sub-communicators, each of which performs an FHI-aims calculation concurrently. This enables massive parallelization of the steps involving energy evaluations and geometry relaxations. The master-slave paradigm is used to allow efficient and automated management of these tasks. The world root (master) rank is not passed into FHI-aims. Rather, it keeps track of which structures have completed and assigns uninitiated structures to root ranks of sub-communicators, which proceed to run FHI-aims.

\subsection{Unit cell volume estimation}
\label{subsection:Unit cell volume estimation}

\par The solid form volume of a molecule is defined as the volume of the unit cell divided by Z, the number of molecules contained in the unit cell. Accurate prediction of the solid form volume of an input molecule is critical for generating structures with reasonable unit cell volumes. To this end, a machine learned model using a Monte Carlo volume estimation scheme and a topological molecular fingerprint constructed based on atomic neighborhoods was developed. The model was trained on a dataset obtained from the CSD using the Conquest program \cite{conquest}. A chemically diverse dataset was compiled, containing molecules with 5 to 260 atoms comprising the organic elements, H, C, N, O, all the halogens, F, Cl, Br, and I, as well as B, P, S, Si, Te, and Se. The accuracy of the machine learned model is within the range of polymorph density differences as identified from  2,173 unique, homomolecular polymorph pairs from the CSD. 


\subsubsection{Dataset construction}

\par The dataset used for training the volume estimation model was obtained from the CSD using the Conquest program \cite{conquest}. The search was performed over entries of the 2017 version of the CSD for structures of homomolecular organic crystals, characterized at room temperature, under standard pressure, and containing the text phrase `polymorph'. As described elsewhere \cite{van2005searching, cruz2015facts, kersten2018survey}, all polymorphic compounds in the CSD are flagged with the tag `polymorph'. All duplicate structures were identified using the COMPACK program \cite{chisholm2005compack} and removed to prevent bootstrapping of the underlying distribution. This yielded 3,768 individual entries in the dataset and 2,173 unique polymorph pairs, which is similar in size to previous statistical studies of homomolecular polymorphs \cite{cruz2015facts,kersten2018survey}.

\par
\begin{figure}[h]
    \centering
    \includegraphics[width=0.6\textwidth]{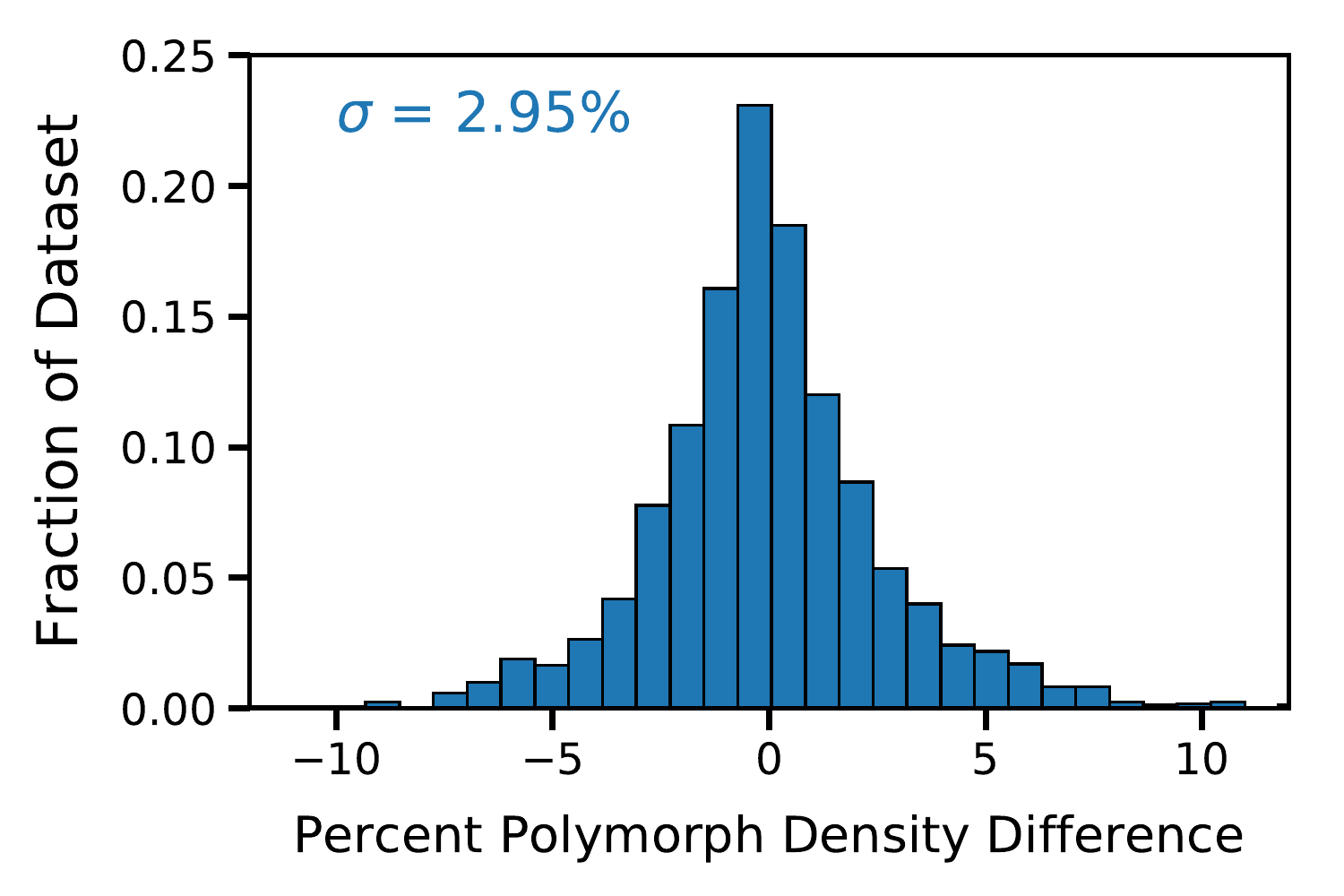}
    \caption{Histogram of the percent difference of polymorph density for 2,173 unique pairs of polymorphs in a dataset obtained from the CSD. The standard deviation of the distribution is displayed in the top left corner.}
    \label{fig:polymorph_volume}
\end{figure}

 The expected variance of the percent difference in the solid form volume of a molecule due to polymorphism was calculated using this dataset. All unique pairs of polymorphs were identified and the percent difference between each polymorph density was calculated. The percent difference of densities is equivalent to that of the solid form molecular volume because the molecular weight remains constant for these systems. The distribution of percent differences is plotted in Figure \ref{fig:polymorph_volume}. The distribution has a standard deviation of 2.95\% with respect to the solid form volume of the molecule, consistent with numerous previous reports of molecular crystal density estimation \cite{burger1979polymorphism, ammon1998new, hofmann2002fast, ye2008new}. This indicates that polymorphs which can exist under the same temperature and pressure conditions could posses significant volume differences owing to the complex nature of the relatively weak intermolecular interactions that govern the lattice energy of homomolecular crystals. Thus, the distribution presented in Figure 2 places a lower bound on the expected accuracy of estimated solid form molecular volumes.

\subsubsection{Monte Carlo volume estimation}

\par Volume estimation is performed by placing a sphere with a van der Waals (vdW) radius \cite{bondi1964van} at the position of each atom in the molecule \cite{gavezzotti1983volumes,pacios1994arvomol}. A Monte Carlo method is then used evaluate the volume occupied by the spheres. First, a three-dimensional box encompassing the molecule is defined. Points within the box are sampled randomly and determined if they fall within at least one of the atomic vdW spheres. The ratio between the number of sampled points and the number of points found within a sphere multiplied by the volume of the three-dimensional box is the estimated volume of the molecule. The Monte Carlo volume estimation is deemed to be converged when the estimated volume changes by less than $10^{-3}$ after $10^6$ new points are samples. 

\par The ratio between the experimental molecular solid form volume and the Monte Carlo volume estimate for the polymorph dataset was found to be 1.47, indicating that the Monte Carlo method systematically underestimates the true solid form volume. Using this linear relationship to predict the solid form volume of the molecule achieves a standard deviation of 4.72\% error with respect to the dataset (Figure \ref{fig:model_performance}). To improve the accuracy of the volume estimation model, specific information about the chemical environment of the atoms in the molecule must be included. To this end, a molecular topological fragment representation has been developed.

\subsubsection{Molecular topological fragment model}

\par We present a topological molecular fingerprint representation for predicting solid form molecular volume within the accuracy of polymorph density differences. The representation is based on molecular fragments determined through analysis of the CSD dataset. The fact that the fragments are not predefined enables an unbiased choice of fragments such that they can represent any structural class. The complexity of the model increases with the size and chemical diversity of the dataset making this representation amenable to large datasets as well as datasets comprising a restricted chemical space. Moreover, representation is invariant to permutations of the atom indexing. The molecular topological fragment representation can be used to predict any molecular property of interest with linear and non-linear regression or classification models and can also be used to compute chemical similarity between molecules using metrics such as the Tanimoto coefficient \cite{maggiora2013molecular}. The Genarris 2.0 source code includes a model construction Python class, enabling users to quickly build topological fingerprints for a training dataset, regularize the model, evaluate the accuracy on a target dataset, and output graphs of predicted values versus target values to asses the performance of the model.

\par The molecular topological fragment representation is built by constructing a unique string representation for every atom in the molecule. Given an atomic environment, the string is deterministic and does not coincide with another distinct atomic neighborhood. The bonding of the molecule is calculated and used to construct a graph comprised of nodes and edges corresponding to atoms and bonds. Then, the atomic neighborhood strings are constructed for every atom in the molecule. The atom's bonded neighbors are identified and concatenated into a string. The string is sorted first by terminal elements in alphabetical order, then by the atom itself if it is non-terminal, and finally the elements the atom is bonded to are given in alphabetical order. All unique neighborhoods across the dataset are collected and sorted in alphabetical order. This ordering is used to index the vector representation of the molecules. For a given molecule, the value of the vector at each index corresponds to the number of each type of fragments present in the molecule. An example of vector representations for glycine and benzene is seen in Table 1. The representation described here is similar to other fragment based representations seen in chemical informatics \cite{rogers2010extended, hansen2015machine}. 

\vspace{-0.5 cm}
\begin{center}
\captionof{table}{Example of vector representations constructed for a dataset containing benzene and glycine using the molecular topological fragment model.}
\begin{tabular}{| c c c c c c c c |}
    \hline
     Fragment & HC & HCCC & HHCCN & HHHNC & HN & OC & OOCC \\ \hline
     Benzene & 6 & 6 & 0 & 0 & 0 & 0 & 0 \\
     Glycine & 2 & 0 & 1 & 1 & 3 & 2 & 1 \\ \hline
\end{tabular}{}
\end{center}{}

\par To construct a predictive model for solid form molecular volumes, the volume predicted by the Monte Carlo method was concatenated to the topological fragment representation vector of each molecule.  The coefficients for a linear model were then calculated using Bayesian ridge regression as implemented in scikit-learn \cite{pedregosa2011scikit}. The regularization parameter was optimized using a grid search method and five-fold cross validation. The number of features contained in the model was constrained by removing features that did not occur at least thirty times in the dataset. Thirty was identified as the optimal number using a five-fold cross validation scheme. This left 64 unique molecular fragments in the model. 

\par
\begin{figure}[h]
    \centering
    \includegraphics[width=0.6\textwidth]{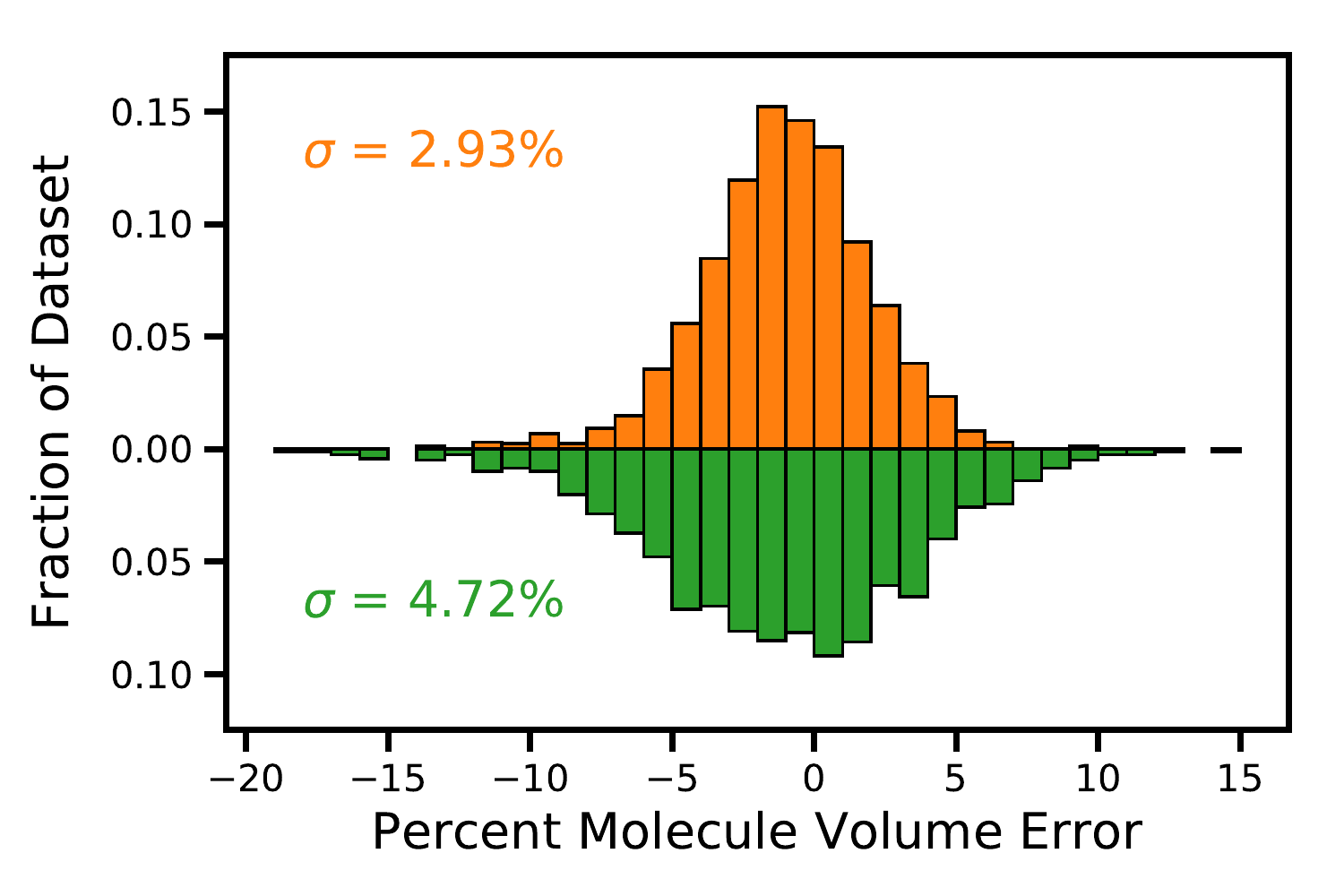}
    \caption{Percent error of the predicted solid form volume for the described dataset from the CSD for a linear model using Monte Carlo volumes and a linear model using Monte Carlo volumes in green and the topological fragment representation of the molecules in orange.}
    \label{fig:model_performance}
\end{figure}

 The distribution of errors obtained using the topological fragment model is displayed in Figure \ref{fig:model_performance}. It is shown that the fragment based model significantly reduces the error in the predicted solid state molecular volumes compared to the Monte Carlo volume estimation. Furthermore, the fragment based model achieves an error of similar magnitude to the volume differences between polymorphs found in the CSD. Thus, the topological fragment model developed here achieves an accuracy within the error one could expect from polymorph density differences. 

\subsection{Structure generation} 
\label{subsection:Structure generation}

The generation process begins by identifying all space groups compatible with the number of molecules in the unit cell. This task is easy for general Wyckoff positions whose multiplicity  must equal the desired number of molecules per cell.  To determine whether a molecule can occupy special Wyckoff positions, its symmetry must be considered.  Genarris 2.0 detects compatible space groups automatically within a given numerical tolerance.  Once the compatible space groups are found, Genarris 2.0 attempts generation of crystal structures sequentially, starting from the lowest space group number. If Genarris is unable to generate a structure within the maximum attempt limit specified by the user, then it proceeds to the next space group.

A random volume is drawn from a Gaussian distribution whose mean and standard deviation are the predicted volume and three times the prediction error of our volume estimation method (see Sec.~\ref{subsection:Unit cell volume estimation}).  The volume is redrawn after a successful generation or after a user-specified number of failed attempts. Subsequently, using this volume,  a unit cell of the desired lattice system is constructed randomly as shown in Figure  \ref{fig:special_position_generation}. If the attempted position is a general Wyckoff position, then the molecule's orientation is sampled randomly and placed randomly inside the unit cell. The space group symmetry operations  are then applied to generate the remaining molecules in the unit cell. Special positions, with the exception of inversion centers,  require alignment of the molecule and their treatment is described in Sec.~\ref{subsubsection:Generation in Special Positions of Space Groups}. The attempted structures that pass the intermolecular distance checks, as described in Sec.~\ref{subsubsection:Structure checks}, are added to the raw pool.

\subsubsection{Generation in special positions of space groups}
\label{subsubsection:Generation in Special Positions of Space Groups}

\begin{figure}
    \centering
    \includegraphics[width=1\textwidth]{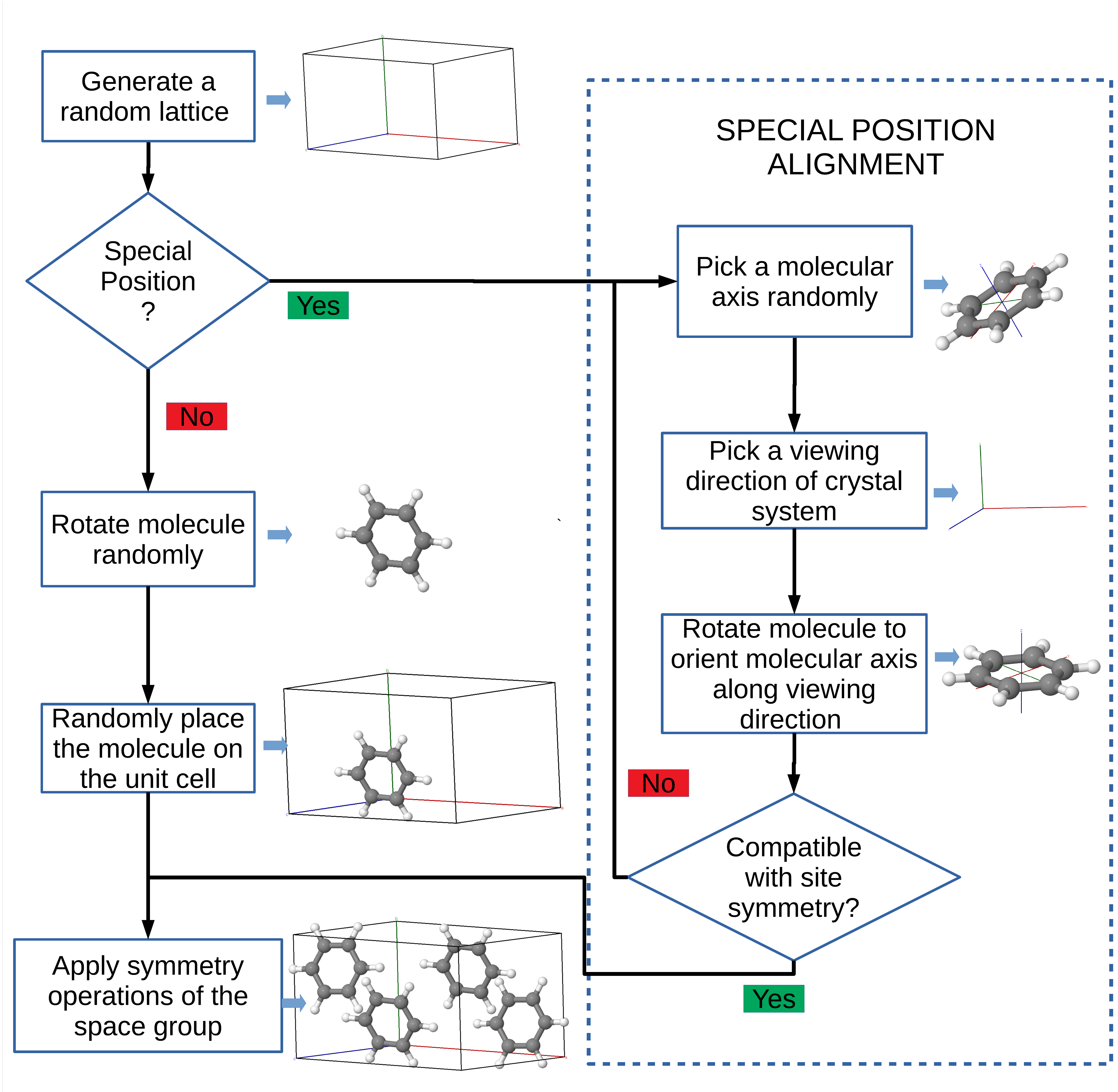}
    \caption{Flowchart of crystal structure generation in Genarris 2.0. Molecules are placed in general Wyckoff positions with a random orientation. In contrast, special positions require specific orientations of the molecule to be compatible with the site symmetry}
    \label{fig:special_position_generation}
\end{figure}

Special Wyckoff positions are left invariant under at least one symmetry operation of the space group in addition to the identity operation. For each space group, the International Table of Crystallography \cite{hahn1983international} lists the special positions whose multiplicity is lower than that of the general position. Only molecules with appropriate symmetry can occupy a special position. Since most molecules do not have higher order symmetries, molecular crystals with molecules on special positions are infrequent. According to an analysis of the CSD \cite{pidcock2003database},  in 70.1\% of the molecular crystals, molecules occupy general positions, and in the remaining structures molecules occupy special positions. Among the special positions, two-fold rotation ($2$), mirror planes ($m$), and inversion centers ($\bar{1}$) are the most frequent. 

Genarris 2.0 generates molecular crystals with molecules on special positions by checking all possible orientations of the molecule with respect to the symmetry directions of the crystal system \cite{hahn1983international}, as shown in the flowchart in Figure \ref{fig:special_position_generation}. At the start of generation, the program finds all possible molecular axes that may be associated with a symmetry element. For this purpose, first the center of mass of the molecule is shifted to the origin. Then, all atoms of the same element that are farthest from the center of mass are selected. The possible symmetry elements of the molecule would map any of these atoms onto itself or onto another. The axes corresponding to these symmetry elements are obtained by calculating the averages and cross products of the position vectors of the selected atoms. A list of potential molecular axes is constructed. To keep the length of the list minimal, parallel vectors are deleted.

Once the potential molecular axes are identified, the code proceeds to check the compatibility of molecule placement at a special position with the specified number of molecules in a unit cell. The molecule's center of mass is placed in a special position such that one of the molecular axes is oriented along one of the viewing directions of the crystal system. Then, the symmetry operations of the space group are applied. If the number of overlapping molecules generated is equal to the order of the site, the special Wyckoff position is regarded as compatible. If not, different molecular axes and viewing directions are considered. All combinations of molecular symmetry axes and viewing directions are examined and compatible ones are stored for subsequent generation attempts. Once a molecule is placed in a compatible special position, its geometry is slightly adjusted (within a user-defined tolerance) by averaging over the atomic positions of all the overlapping molecules occupying the same site. Depending on the site symmetry of the special position, the allowed degrees of freedom are randomized. For example, on a general position or an inversion center, the molecule can be freely rotated about its center of mass. A molecule placed on a two-fold axis can be freely rotated about the axis. 

\subsection{Structure checks}
\label{subsubsection:Structure checks}

Attempted structures are checked to avoid unphysically close intermolecular contacts. Checking the distance between every atom of a molecule and every atom of all neighboring molecules, including its own periodic replicas, has a scaling of $O(N^2)$, where $N$ is the number of atoms in the unit cell. This is found to be the bottleneck of structure generation. To improve the efficiency, Genarris 2.0 performs a series of three hierarchical structure checks. Failed structures are discarded at each stage, such that fewer structures undergo the more rigorous and computationally expensive checks.

The threshold for allowed close contacts between two atoms is called the cutoff distance and is defined based on a specific radius fraction, $s_r$, of the sum of atomic van der Waals radii \cite{li2018genarris}. The crystal structure is deemed unphysical and rejected if the distance, $d$, between two atoms belonging to different molecules is such that
\begin{equation}
d < s_r(r_A+r_B)
\end{equation} 
where $r_A$ and $r_B$ are the van der Waals radii of atom A and atom B, respectively. This ensures the quality of the generated structures. The default value of $s_r$ is $0.85$. Based on statistical analysis of structures extracted from the CSD, this is a reasonable setting for all but the strong hydrogen bonds. For these cases, special settings have been implemented in Genarris 2.0, as described in Sec.~\ref{Hbonds}.
For this value of $s_r$ and the target unit cell volume determined as described in Sec.~\ref{subsection:Unit cell volume estimation}, random generation of crystal structures may require a large number of attempts (a few thousand to millions) before it passes all three stages of structure checks and is accepted into the pool. Therefore, the new hierarchical structure check procedure is a significant efficiency improvement in Genarris 2.0. The details of each stage are explained below.

\subsubsection{Stage I: Fast screening without periodic boundary conditions}

 For preliminary screening, periodic boundary conditions are completely ignored and only intermolecular distances in the unit cell are evaluated. Because distances are computed using the Euclidean norm,  this stage is the fastest. If the centers of mass of a pair of molecules in a cell are much farther than twice the molecule length, defined as the maximum distance between two atoms of a molecule, then those pairs are ignored as these molecules cannot overlap. We find that most of the unphysical structures generated are rejected at this stage. The structures that pass this screening proceed to the second stage of structure checks.

\begin{figure}
    \centering
    \includegraphics[width=0.9\textwidth]{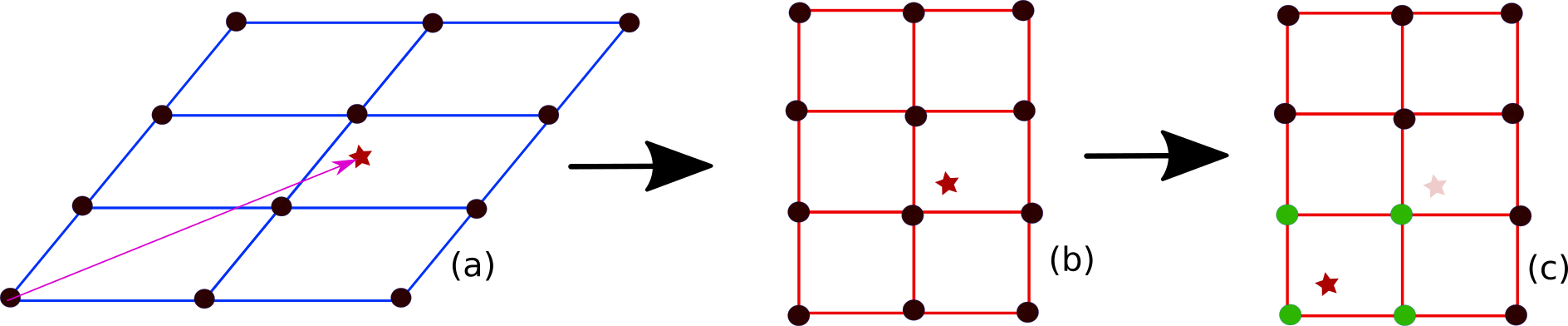}
    \caption{A two dimensional representation of Stage II approximate distance evaluation under periodic boundary conditions. a) An oblique lattice in Cartesian coordinates. The star denotes the point whose distance to the nearest lattice point we need to find. b) Once the lattice is converted from real space into the fractional basis, it is easy to find the box that bounds the point. c) The nearest lattice point is likely to be one of the real space points that map to the green points.}
    \label{fig:tier2}
\end{figure}

\subsubsection{Stage II: Distance checks with periodic replicas}

 In this step, the distances of a molecule from other molecules in the unit cell as well as its own periodic images are checked against the cutoff distance. An approximate minimum image convention is implemented for non-orthogonal cells. To accelerate the distance checks, non-orthogonal cells undergo a lattice reduction. Let \textbf{a} = $[a_x, 0, 0]$, \textbf{b} = $[b_x, b_y, 0]$, and \textbf{c} = $[c_x, c_y, c_z]$ be the lattice vectors in a Cartesian coordinate system. It is possible to choose a less oblique lattice which satisfies: $a_x, b_y ,c_z > 0$;
$|b_x|, |c_x|\leq a_x/2$; and $|c_y|\leq b_y/2$.

The Stage II algorithm is illustrated in Figure \ref{fig:tier2}. First, the atom positions are expressed in fractional coordinates. Then, the distance between two atoms is computed in fractional space and translated to the ``origin cube", spanned by the vectors $[1,0,0], [0,1,0], [0,0,1]$. Finally, the minimum Cartesian distance of this point from the corners of the origin cube is calculated. For orthogonal cells, the closest point in fractional space necessarily corresponds to the closest point in real Cartesian space. However, for an oblique triclinic lattice a different lattice point may be closer to this point. Therefore, if a non-orthogonal structure passes Stage II, it proceeds to Stage III for a more rigorous check.

\subsubsection{Stage III: Rigorous checks for non-orthogonal cells}

Complete structure checks require exact evaluation of distances under minimum image convention. For non-orthogonal cells, this problem is a three-dimensional case of  the well-studied closest vector problem \cite{cvp2002}. If the lattice is translated such that one of the two points coincides with the origin, we need to find the distance $d$ to the nearest lattice point of the position vector $\mathbf{x}$ of the second point. That is, 

\begin{equation}
 d^2= \min_\mathbf{n}  | \mathbf{L}^T \mathbf{n}- \mathbf{x}|^2 ,
\end{equation}
where $\mathbf{n} = [n_x, n_y ,n_z]$; $n_x, n_y ,n_z \in \mathbb{Z}^3$; and $\mathbf{L}=[\mathbf{a}, \mathbf{b}, \mathbf{c}]^T$. This problem is encountered in communication theory, where the received signal over a communication line is decoded by finding the nearest lattice point \cite{rogers2016overcoming}. One popular approach is the Finck and Pohst sphere decoder \cite{fincke1985improved, hassibi2005sphere} method, where the closest lattice point is found using a tree search and the depth of the tree corresponds to the dimension of the problem. Genarris 2.0 uses a version of the sphere decoder to compute the exact distance under minimum image convention for non-orthogonal cells. The distance estimate obtained from Stage II is used as the initial sphere radius for the sphere decoder algorithm. This step is the slowest, but only few non-orthogonal structures that pass Stage I and Stage II reach Stage III. Hence, the overall efficiency is not compromised.

\subsubsection{Intermolecular cutoff distances}
\label{Hbonds}

\par
\begin{figure}[h]
    \centering
    \includegraphics[width=0.8\textwidth]{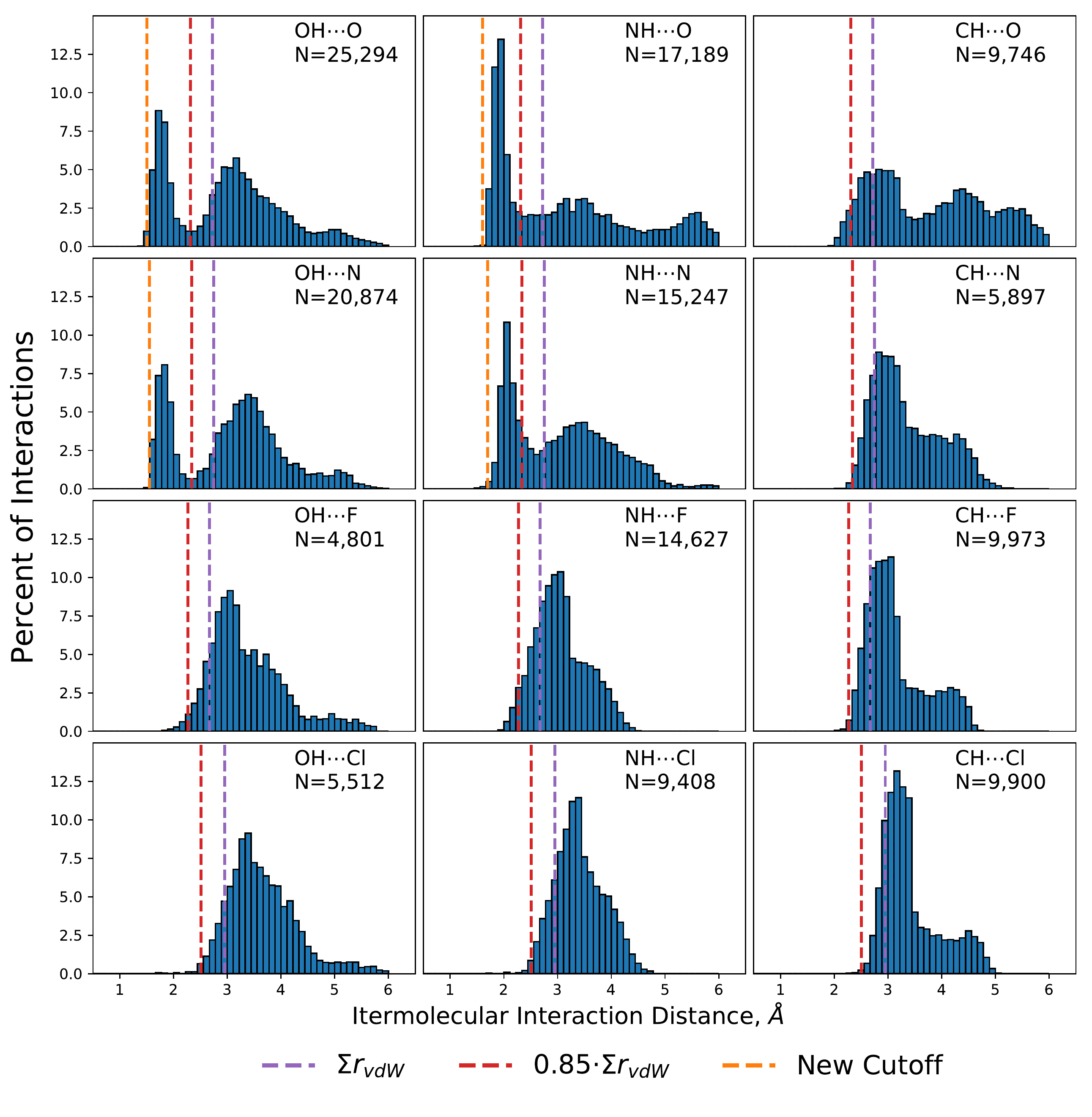}
    \caption{Plots of the number of observations as a function of distance for intermolecular contact distances mined from the CSD and gathered from the IsoStar program. Each histogram is labeled with its hydrogen bond and the number of interactions obtained from IsoStar are labeled N. Drawn on the histograms are vertical lines at the distance corresponding to the sum of the vdW radii, a vdW cutoff of 0.85, and the new hydrogen bond cutoff distances.}
    \label{fig:h_bond}
\end{figure}

 Choosing appropriate intermolecular cutoff distances is critical for generating physically reasonable structures. In Genarris 2.0, cutoff distances are a function of the elements participating in the intermolecular interaction. For vdW interactions, cutoff distances are implemented using an $s_r$ of 0.85. An $s_r$ of 0.85 was determined to be a physically reasonable value based on our statistical analysis of intermolecular contacts in a data extracted from CSD and presented in Figure \ref{fig:h_bond} as well as an earlier analysis \cite{rowland1996intermolecular}. However, for hydrogen bonds, the intermolecular distance may be considerably shorter than the $s_r$ value used for weaker intermolecular interactions \cite{rowland1996intermolecular, steiner2002hydrogen}. Hence, new settings for the allowed interatomic distances for hydrogen bonds have been implemented in Genarris 2.0.

\par Hydrogen bonds among the most important intermolecular interactions in both naturally occurring and artificially engineered molecular crystals \cite{steiner2002hydrogen}. Intermolecular hydrogen bonds are denoted as XH$\cdots$Y where X is the donor, which is covalently bonded to the hydrogen, and Y is the acceptor, which belongs to a different molecule than X. The cutoff distance between H and Y implemented in Genarris 2.0 depends on the identity of atoms X and Y. However, these cutoff distances are applicable to any functional group pair that would participate in an intermolecular hydrogen bond for homomolecular crystals. 

\pagebreak
\begin{center}

\captionof{table}{New cutoff distances implemented in Genarris for intermolecular hydrogen bonds. The cutoff distances are compared to the sum of the van der Waals radii for the intermolecular interactions using the specific radius ($s_r$) fraction defined in Sec.~\ref{subsubsection:Structure checks}.}
\begin{tabular}{|c| p{45mm} | p{35mm} | c |}
    \hline
     \textbf{Contact Type} & \centering  \textbf{Cutoff Distance} ($\AA^3$) & \centering  \textbf{Sum of van der Waals radii} ($\AA^3$) & \textbf{$s_r$} \\ \hline
     OH$\cdots$O & \centering 1.5 & \centering  2.72 & 0.55 \\
     OH$\cdots$N & \centering  1.6 & \centering  2.75 & 0.58 \\
     NH$\cdots$O & \centering  1.6 & \centering  2.72 & 0.59 \\
     NH$\cdots$N & \centering  1.7 & \centering  2.75 & 0.62 \\
     \hline
\end{tabular}
\end{center}

\par Table 2 displays the newly implemented contact distances for hydrogen bonds, in which oxygen or nitrogen are the donor and acceptor. These values were determined based on the existing literature \cite{gilli2009nature}, as well as statistical searches of the CSD using the IsoStar program \cite{bruno1997isostar}. The IsoStar program provides distributions of nonbonded, intermolecular distances between pairs of functional groups. The central and contact functional groups were chosen across the available pK$_a$ range \cite{gilli2008predicting} for each type of hydrogen bond in order to develop general three body cutoff distances for all relevant hydrogen bonds. The results of the IsoStar searches are shown in Figure \ref{fig:h_bond}. For hydrogen bonds involving oxygen and nitrogen as the donor and acceptor, the sum of the vdW radii multiplied by the default $s_r$ value of 0.85 (red dashed lines) exceeds a large number of non-bonded interaction distances, illustrated by the left-most peak of the bimodal distributions. Using the default $s_r$ value, structures with strong hydrogen bonds, such as glycine, would be deemed unphysical and discarded. With the new settings listed in Table 2, they would be considered physically reasonable. For hydrogen bonds involving halogens \cite{brammer2001understanding}, or those with carbon as the donor atom \cite{steiner1996c}, the default $s_r$ value of 0.85 is still appropriate. 


\subsection{Clustering and down-selection}
\label{subsection:Clustering and down-selection}

\begin{figure}[h]
    \centering
    \includegraphics[width=0.23\textwidth]{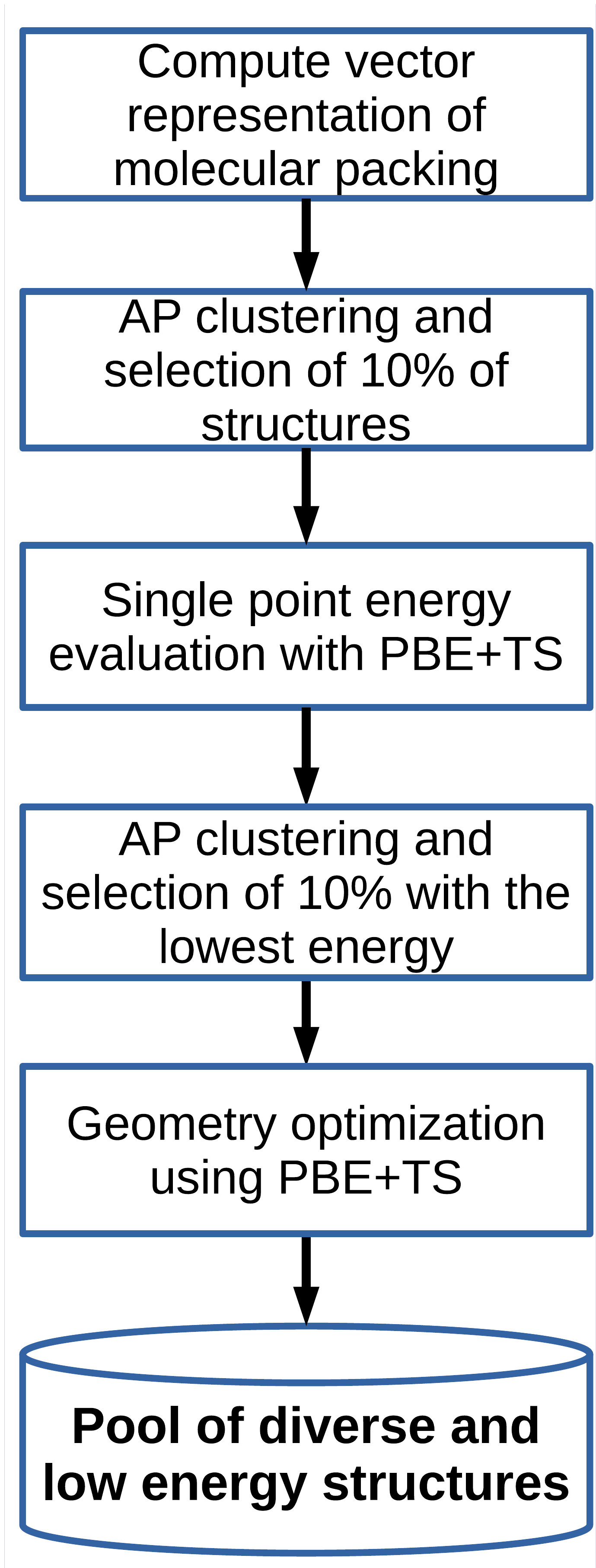}
    \caption{The clustering and down-selection steps of the ``Robust" workflow of Genarris 2.0.}
    \label{fig:Clustering}
\end{figure}

Once a ``raw" pool of physically reasonable, random structures is generated, Genarris 2.0 offers the option of performing a user-defined sequence of clustering, energy evaluation, and down selection steps in order to form a smaller curated pool of structures. Here, we use the Robust workflow, as shown in Figure~\ref{fig:Clustering}.
For the purpose of clustering via the affinity propagation (AP) machine learning algorithm \cite{frey2007clustering}, a feature vector describing the molecular packing is calculated for each structure. The relative coordinate descriptor (RCD) \cite{li2018genarris} and a radial distribution function (RDF) \cite{RDF_1,RDF_2,RDF_3} descriptor are implemented in Genarris 2.0. The \textit{preference} hyperparameter of AP is automatically tuned by Genarris 2.0 to produce the desired number of clusters within a user-defined tolerance, as described in Sec.~\ref{subsection:Parallelization}. The default number of clusters for this step is $10\%$ of the the number of structures in the raw pool. For the exemplar of each cluster, a single point energy (SPE) evaluation is performed using FHI-aims with the settings described  in Sec.~\ref{section:DFT settings} below. Then, AP clustering is performed again with the target number of clusters set to 10\% of the reduced pool and the lowest energy structure is selected out of each cluster. Finally, the remaining structures are fully relaxed with FHI-aims as described in Sec.~\ref{section:DFT settings}. This constitutes the final pool of structures output by Genarris 2.0 using the Robust workflow.

\section{DFT settings}
\label{section:DFT settings}
Genarris 2.0 interfaces with the FHI-aims electronic structure code \cite{blum2009ab} for geometry relaxation of the single molecule and of the structures in the final pool, as well as for single point energy (SPE) evaluations within the Robust workflow used here. All invocations of FHI-aims in this work used the Perdew-Burke-Ernzerhof (PBE) generalized gradient approximation \cite{perdew1996generalized} and the Tkatchenko-Scheffler (TS) pairwise dispersion correction \cite{tkatchenko2009accurate} with \textit{lower-level} numerical settings, which correspond to the light/tier1 settings of FHI-aims. SPE calculations for crystals were done self-consistently with a $1\times 1\times 1$ k-point grid for fast screening. Geometry relaxations of the final pool were performed using a $3\times 3\times 3$ k-point grid. Additionally, no constraints were placed on the lattice. All relaxations were done under ambient conditions except for the case of the high-pressure $Z=2$ polymorph of benzene, where the pressure was set to $25$ kbar to reflect the experimental conditions.

\section{Case studies}
\subsection{Benzene}

\begin{figure}[h]
\begin{minipage}{0.48\textwidth}
    
        \centering
        \includegraphics[width=0.9\textwidth]{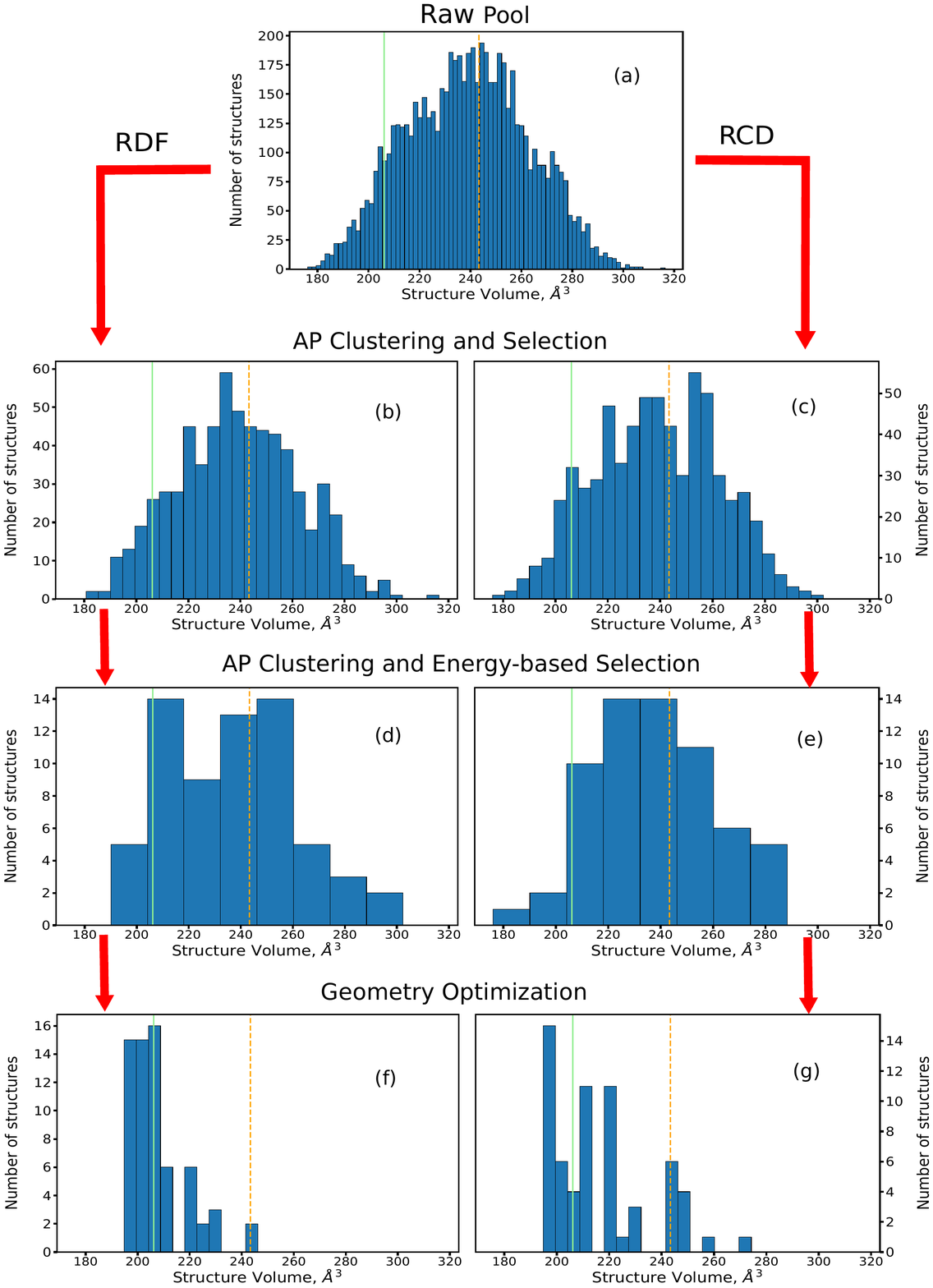}
        \caption{Unit cell volume histograms obtained at each step of the Robust workflow for benzene with $Z=2$.  The solid green line denotes the unit cell volume of the experimental structure and the dashed orange line shows the volume predicted by our model.}
        \label{fig:ben_2mpc_vol}
    
\end{minipage} \hfill
\begin{minipage}{0.48\textwidth}
    
        \centering
        \includegraphics[width=0.865\textwidth]{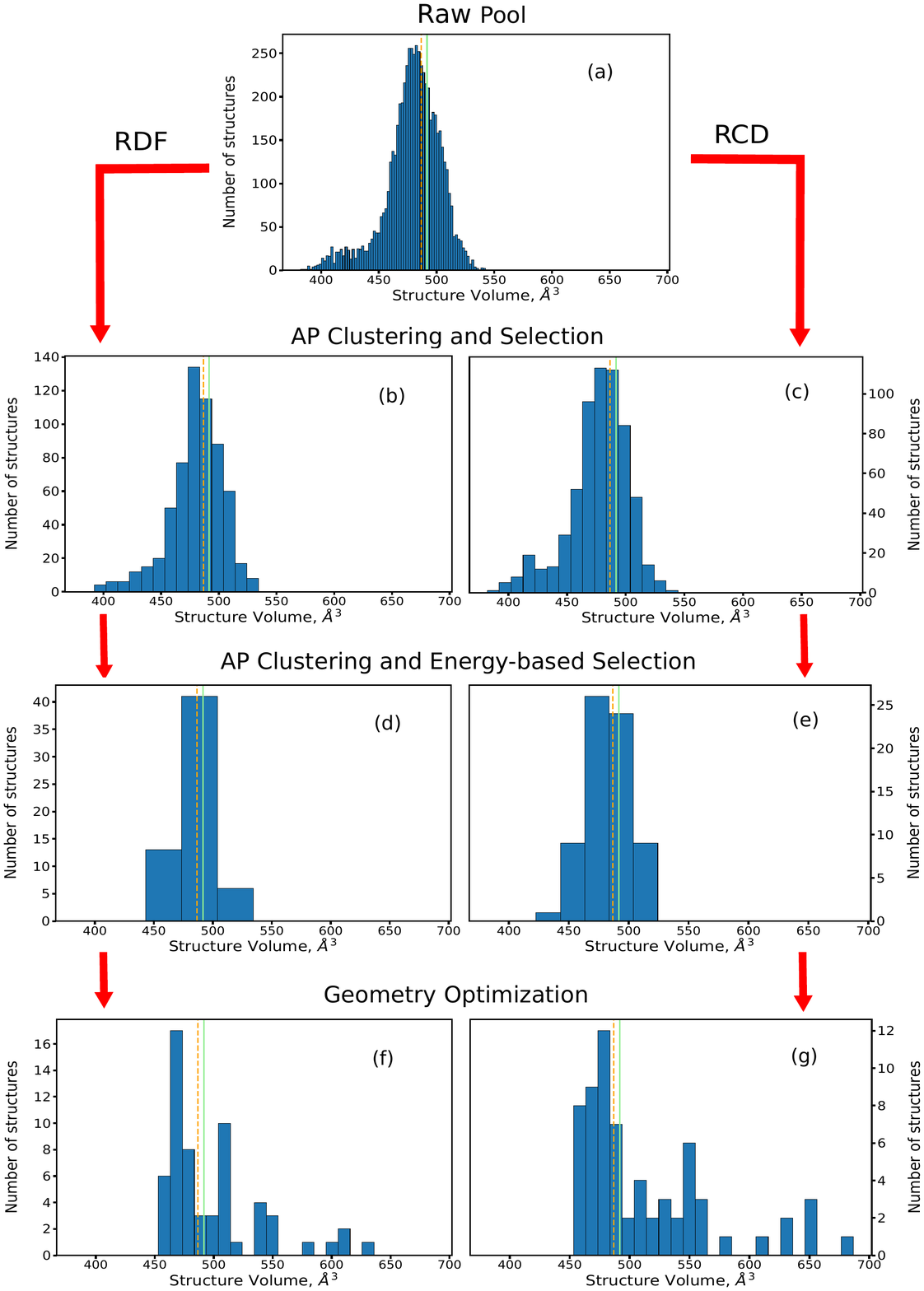}
        \caption{Unit cell volume distributions obtained at each step of the Robust workflow for benzene with $Z=2$. The green line denotes the unit cell volume of the experimental structure and the dashed orange line shows the volume predicted by our model.}
    \label{fig:ben_4mpc_vol}
    
\end{minipage}
\end{figure}


\begin{figure}[h]
\begin{minipage}{0.48\textwidth}
    
   \centering
    \includegraphics[width=0.86\textwidth]{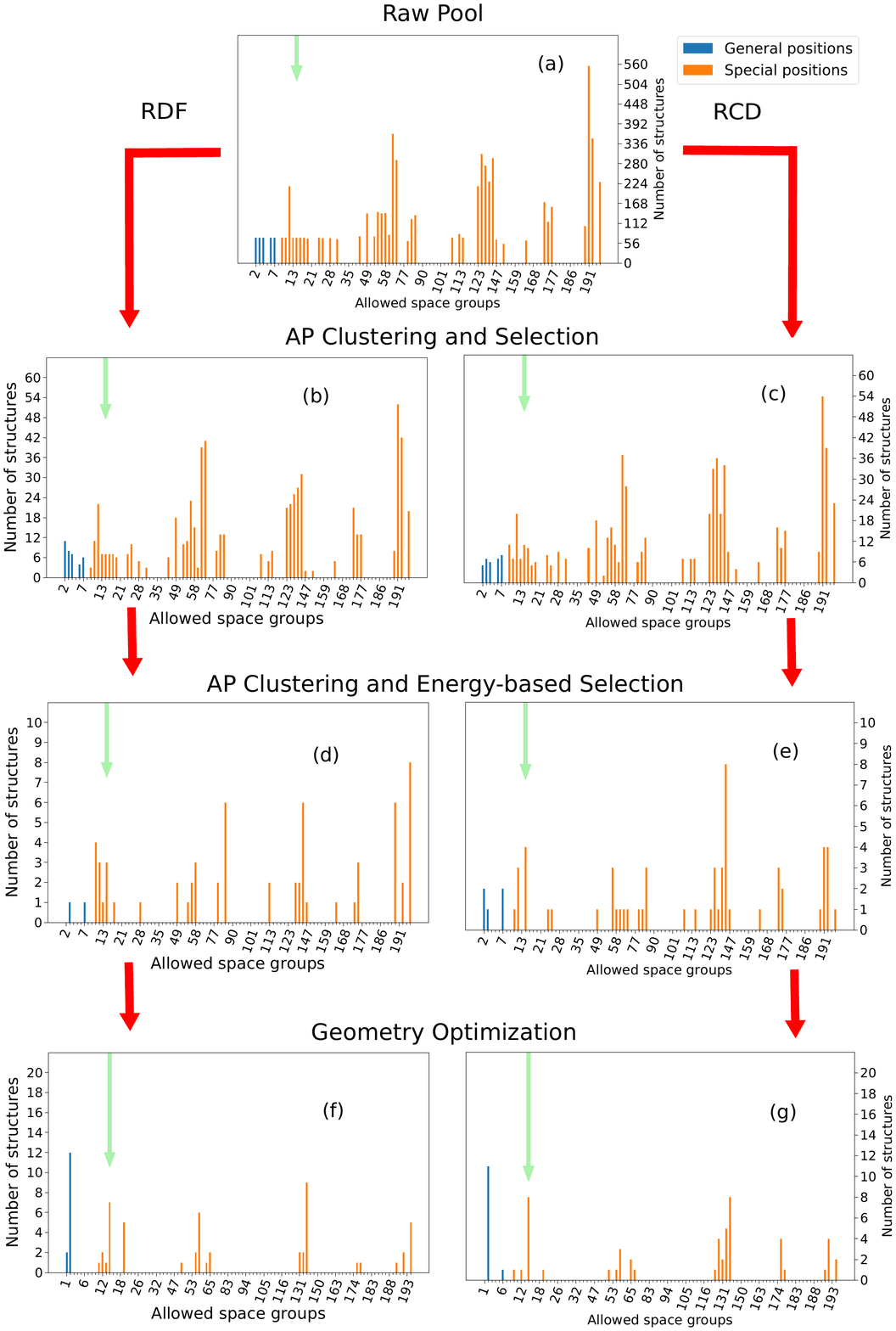}
    \caption{Space group distribution histograms obtained at each step of the Robust workflow for benzene with $Z=2$. The green arrow points to the space group of the experimental structure,  $P2_1/c$ (14)}
    \label{fig:ben_2mpc_spg}

\end{minipage} \hfill
\begin{minipage}{0.48\textwidth}

    \centering
    \includegraphics[width=0.86\textwidth]{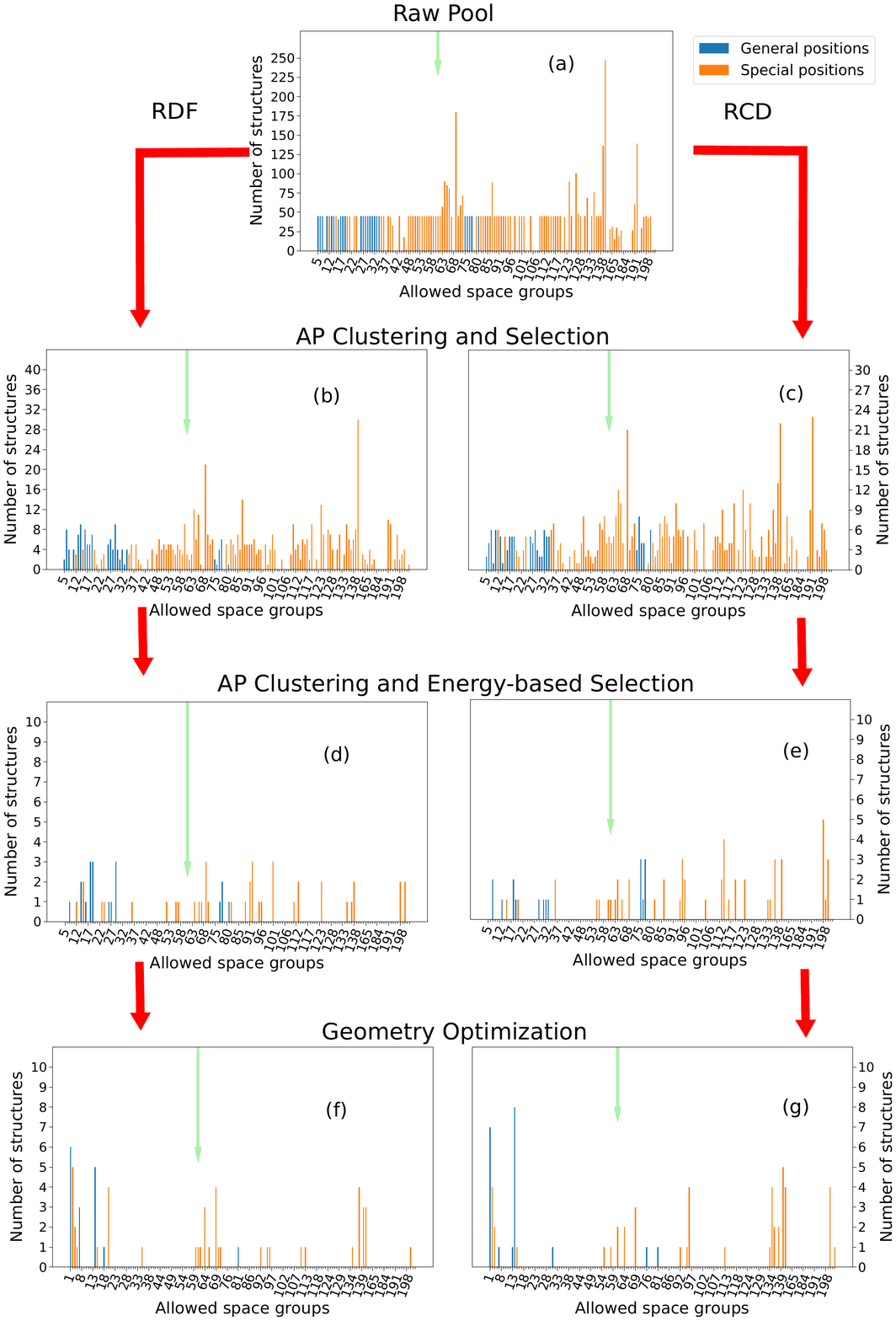}
    \caption{Space group distribution histograms obtained at each step of the Robust workflow for benzene with $Z=4$. The green arrow points to the space group of the experimental structure, $Pbca$ (61).}
    \label{fig:ben_4mpc_spg}

\end{minipage}
\end{figure}

\begin{figure}[h]
\begin{minipage}{0.48\textwidth}

    \centering
    \includegraphics[width=0.85\textwidth]{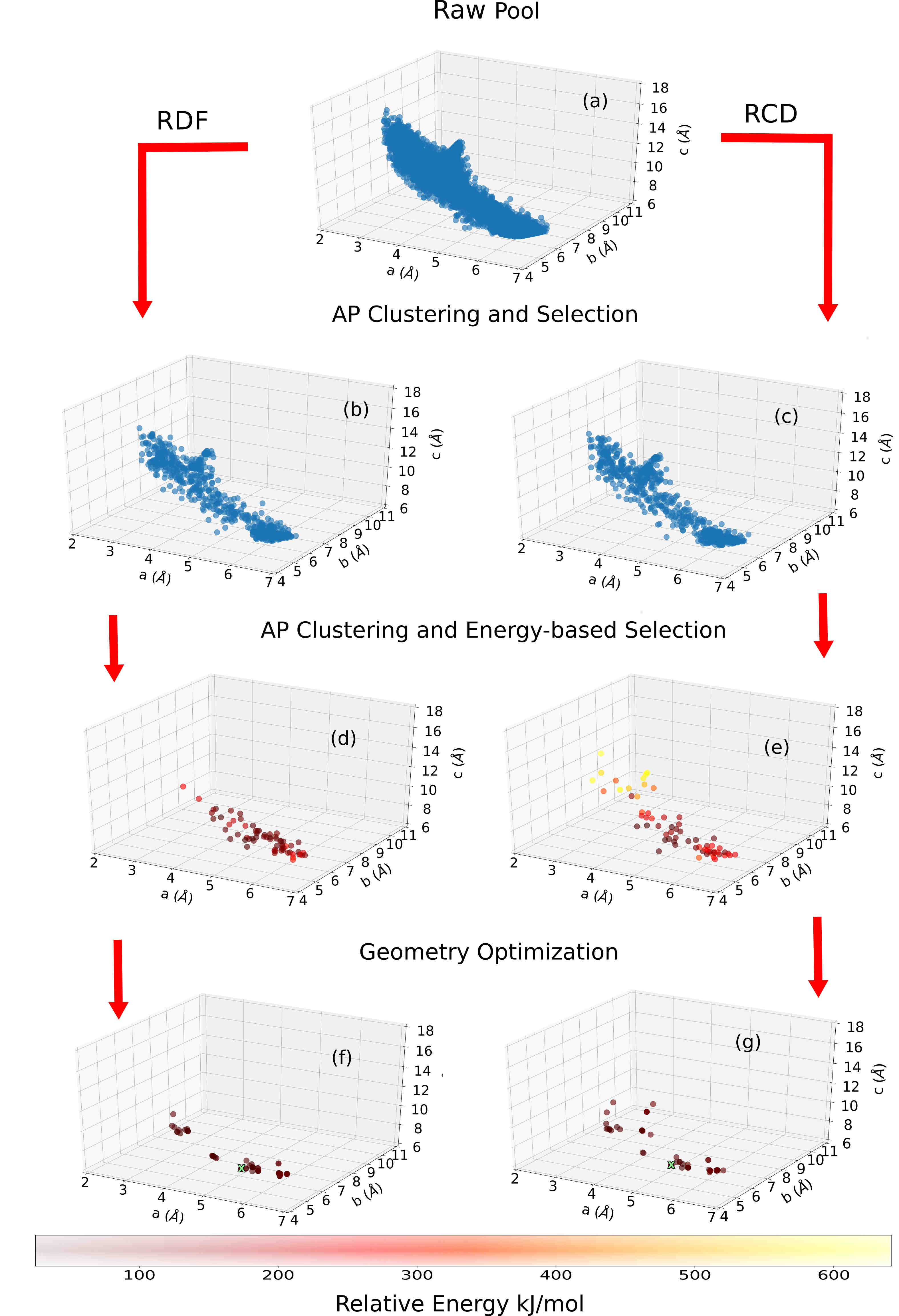}
    \caption{Lattice parameter distributions obtained at each step of the Robust workflow for benzene with $Z=2$. The green cross denotes the experimental structure.}
    \label{fig:ben_2mpc_lat}

\end{minipage} \hfill
\begin{minipage}{0.48\textwidth}

    \centering
    \includegraphics[width=0.85\textwidth]{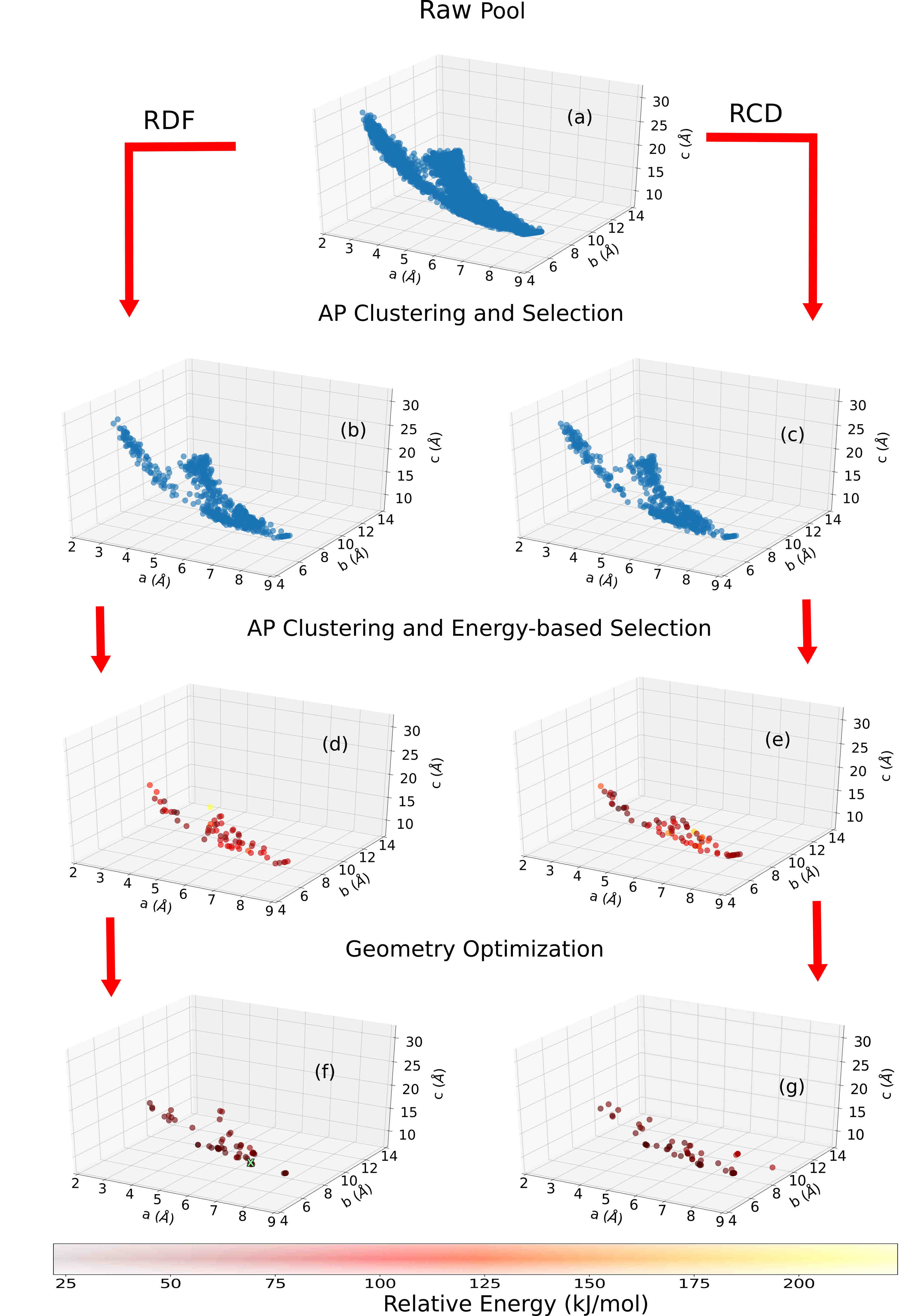}
    \caption{Lattice parameter distributions obtained at each step of the Robust workflow for benzene with $Z=4$. The green cross denotes the experimental structure.}
    \label{fig:ben_4mpc_lat}

\end{minipage}
\end{figure}

With the chemical formula of $C_6H_6$, benzene is one of the simplest aromatic hydrocarbons. It is a highly symmetric molecule with a $6/mmm$ point group which allows special positions with 20 different site symmetries. Two known polymorphs of benzene are \cite{ciabini2007triggering}: a) $Z = 4$ and space group $Pbca$ (61) under ambient pressure and b) $Z = 2$ and space group $P2_1/c$ (14) under high pressure. In both structures, benzene occupies a special position with an inversion center ($\bar{1}$).

Figures \ref{fig:ben_2mpc_vol} and \ref{fig:ben_4mpc_vol} show the volume histograms obtained at each step of the Robust workflow for benzene with $Z=2$ and $Z=4$, respectively. The experimental volume and the predicted volume are indicated by  solid green line and dashed orange line, respectively. Raw pools of about $6000$ structures were generated for both $Z = 2$ and $Z = 4$ with predicted volumes of $243$ $\AA^3$ and $487$ $\AA^3$, and volume standard deviations of $18$ $\AA^3$ and $37$ $\AA^3$, respectively. The volume of the experimental structures were $206$ $\AA^3$ and $471$ $\AA^3$, respectively. Our prediction for $Z=4$ is much closer to the experimental volume than for $Z=2$ because the latter forms under pressure of $25$ kbar whereas the volume estimation model was trained on structures obtained under ambient pressure. Nevertheless, Figure \ref{fig:ben_2mpc_vol} shows noteworthy density about the experimental volume throughout the workflow progression. The resulting volume distributions in Figures \ref{fig:ben_2mpc_vol} and \ref{fig:ben_4mpc_vol} are approximately Gaussian until the relaxation step. For $Z=2$, relaxation under pressure resulted in volume contraction, whereas some $Z=4$ structures expanded beyond the initial volume range. 

Figures \ref{fig:ben_2mpc_spg} and \ref{fig:ben_4mpc_spg} show the space group distributions obtained at each step of the Robust workflow for benzene with $Z=2$ and $Z=4$, respectively. Space groups with general Wyckoff positions are colored in blue and space groups with special Wyckoff positions are colored in orange. Genarris 2.0 attempts to generate a uniform space group distribution. We find that the generated space group distributions are approximately uniform with significant number of structures in the experimental space group for both $Z=2$ and $Z=4$. Some space groups may be very difficult or impossible to generate within the given physical constraints. For example, for $Z=4$, space groups like $P2/m$ (10), $Pmm2$ (25), and $Pmmm$ (47) which have mirror planes are harder to generate as molecules that touch the planes overlap with their own mirror image \cite{pidcock2003database}. In contrast, space groups with glide planes and screws axes are easier to generate because symmetry-equivalent molecules are translated in space. Some structures can have a higher site symmetry on a special position than we attempted to generate, resulting in overpopulation of some space groups. For example, for $Z=2$, space group $P6/mmm$ (191) has a relatively large occupation as shown in panel (a) of Figure \ref{fig:ben_2mpc_spg}. Many of these structures were discarded in the subsequent selection steps. 

Figures \ref{fig:ben_2mpc_lat} and \ref{fig:ben_4mpc_lat} show the lattice parameter distributions obtained at each step of the Robust workflow for benzene with $Z = 2$ and $Z = 4$, respectively. For the energy-based selection and final relaxation steps, the color scale corresponds to relative energies with respect to the lowest energy structure in the final relaxed pool. The lattice parameter distribution of the raw pools resembles the shape of the surface $|a||b||c| =$ constant (an approximate relation given that benzene is able to assume
many lattice types), indicating approximately uniform sampling of the lattice parameter space.  Down-selection based on energy tends to filter out very elongated structures whose $c$ parameter is significantly longer than $a$ and $b$, indicating that these are relatively unstable for benzene. In fact, the experimental unit cells are not elongated. Panels (f) and (g) of Figure \ref{fig:ben_2mpc_lat} shows that relaxation under pressure resulted in a distribution characterized by a few clusters, suggesting that pressure may have restricted the physically feasible regions. For $Z=2$, the experimental structure, indicated by a green X, is found in the final relaxed pools for both the RCD and RDF runs. For $Z=4$, the final pool was more diverse in lattice parameter space as seen in panels (f) and (g) of Figure \ref{fig:ben_4mpc_lat} and the experimental structure was found only in the RDF run. It should be noted that the goal of Robust workflow is not necessarily to find the experimental structure but to adequately sample the configuration space. 

\subsection{Glycine}

\begin{figure} [h]
\begin{minipage}{0.48\textwidth}

    \centering
    \includegraphics[width=0.85\textwidth]{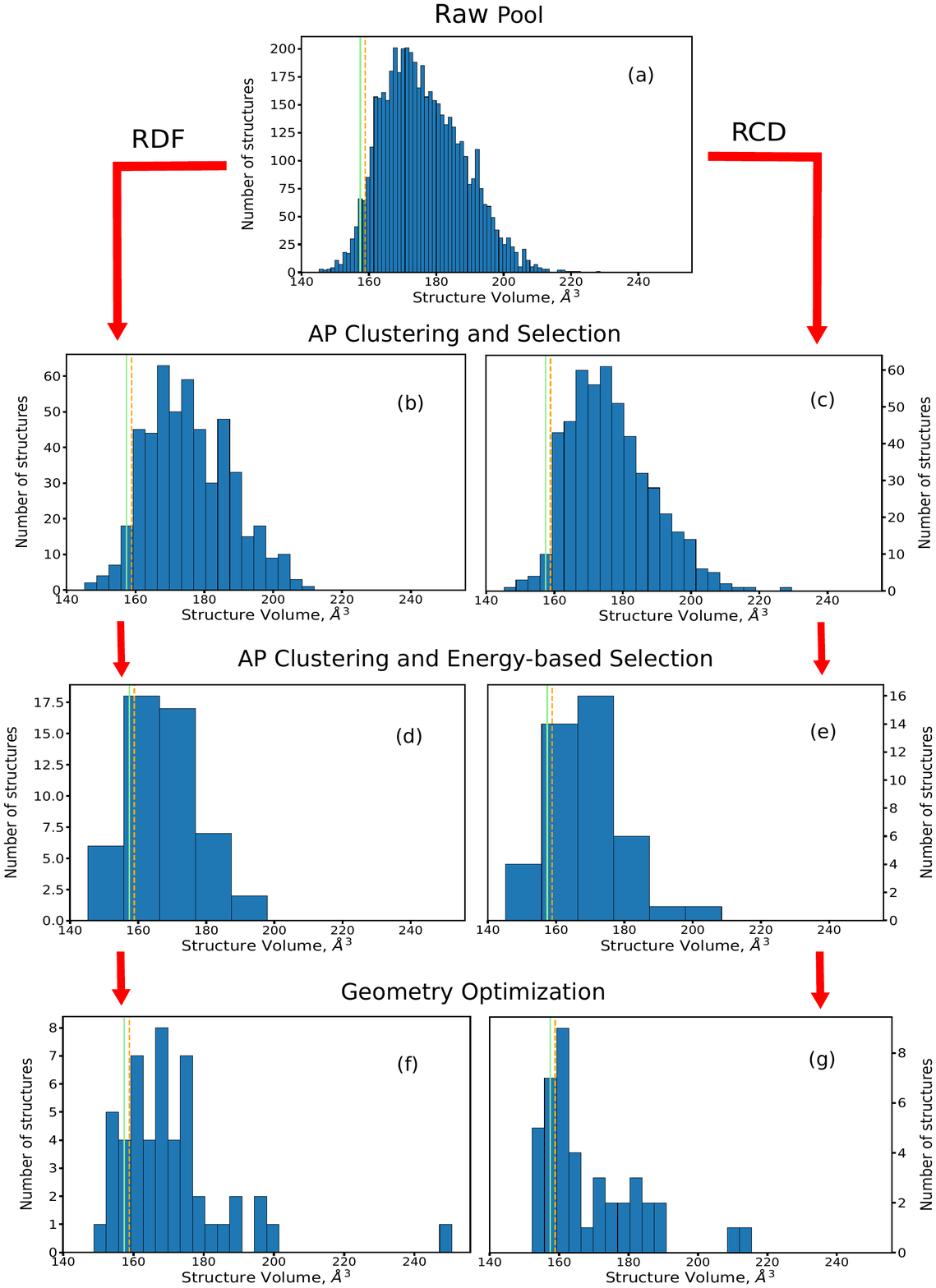}
    \caption{Unit cell volume distributions obtained at each step of the Robust workflow for glycine with $Z=2$. The green line denotes the unit cell volume of the experimental structure and the dashed orange line shows the volume predicted by our model.}
    \label{fig:gly_2mpc_vol}

\end{minipage} \hfill
\begin{minipage}{0.48\textwidth}

    \centering
    \includegraphics[width=0.85\textwidth]{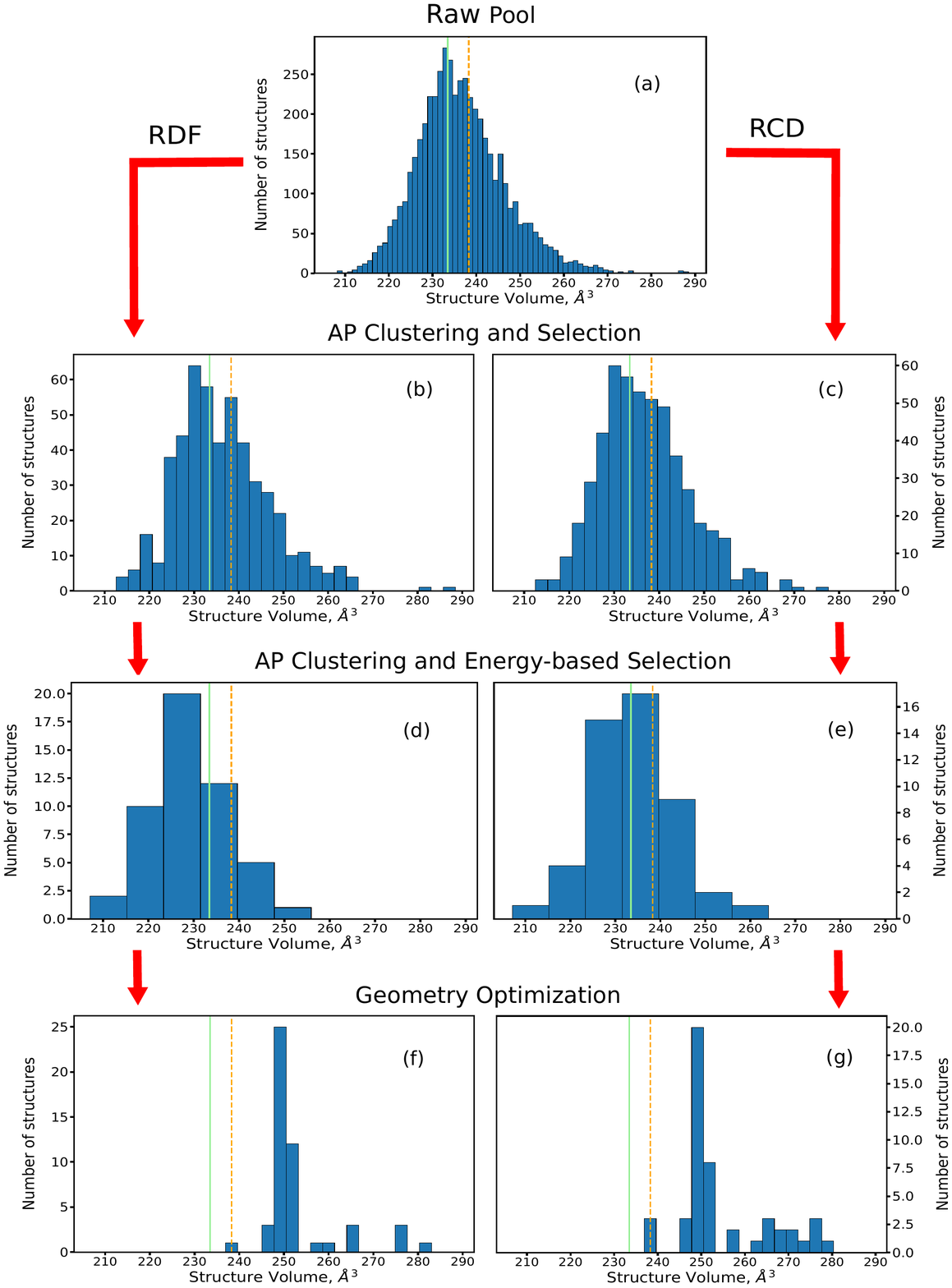}
    \caption{Unit cell volume distributions obtained at each step of the Robust workflow for glycine with $Z=3$. The green line denotes the unit cell volume of the experimental structure and the dashed orange liwith ne shows the volume predicted by our model.}
    \label{fig:gly_3mpc_vol}

\end{minipage}
\end{figure}

\begin{figure}[h]
\begin{minipage}{0.48\textwidth}

    \centering
    \includegraphics[width=0.85\textwidth]{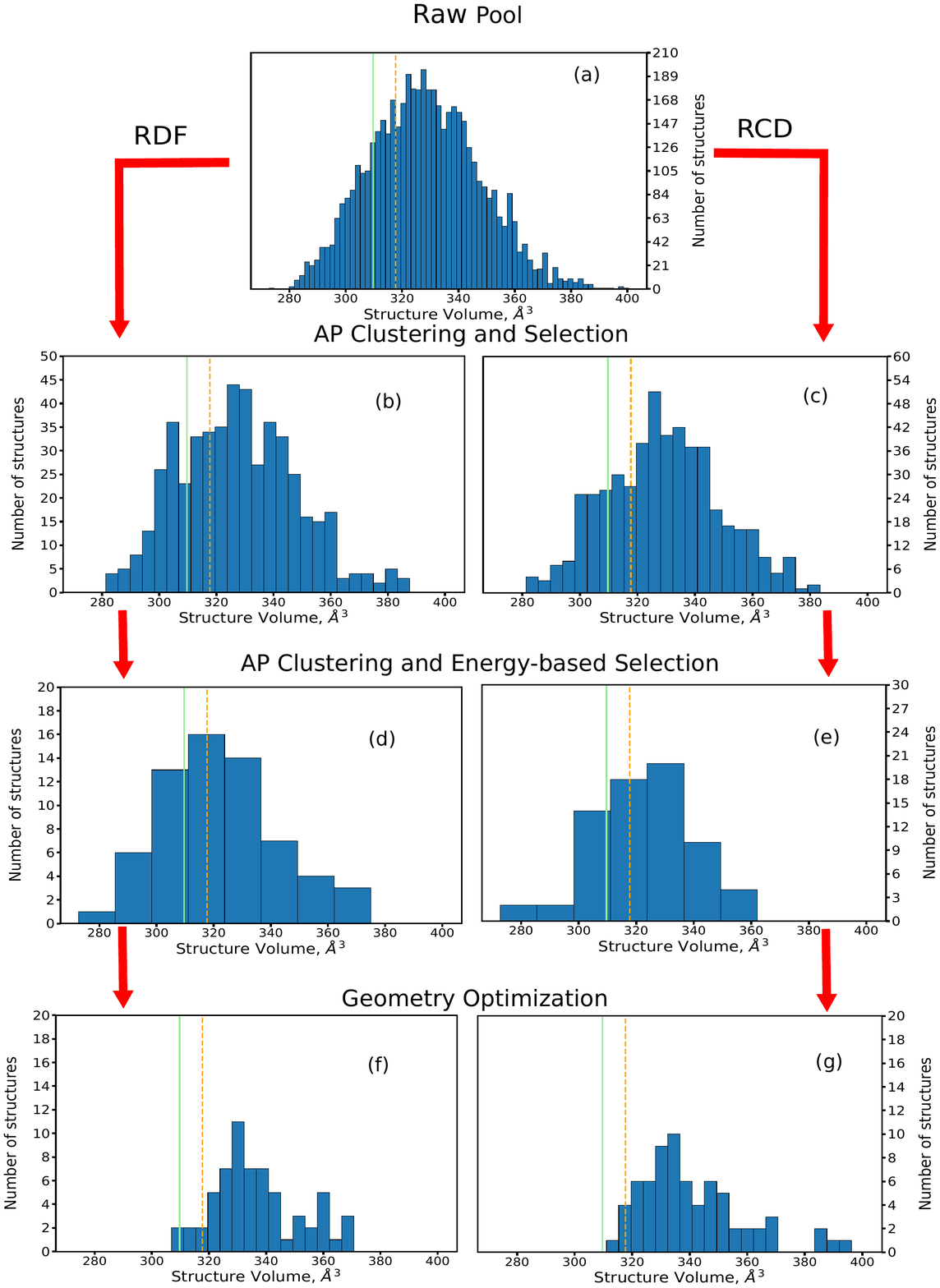}
    \caption{Unit cell volume distributions obtained at each step of the Robust workflow for glycine with $Z=4$.  The green line denotes the unit cell volume of the experimental structure and the dashed orange line shows the volume predicted by our model.}
    \label{fig:gly_4mpc_vol}

\end{minipage} \hfill
\begin{minipage}{0.48\textwidth}

    \centering
    \includegraphics[width=0.85\textwidth]{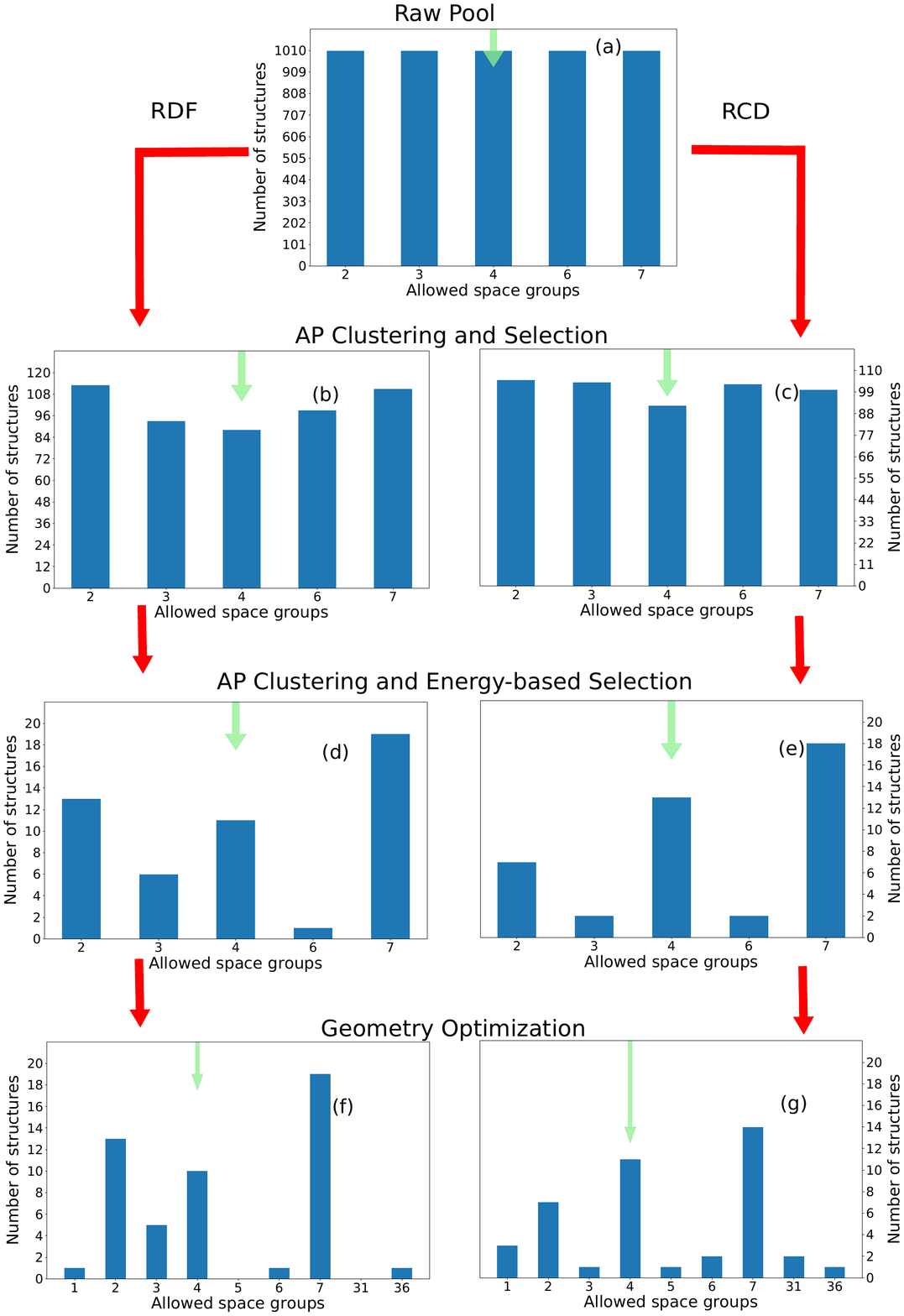}
    \caption{Space group distribution histograms obtained at each step of the Robust workflow histogram for glycine with $Z=2$. The green arrow points to the space group of the experimental structure, $P2_1$ (4).}
    \label{fig:gly_2mpc_spg}

\end{minipage}
\end{figure}

\begin{figure}[h]
\begin{minipage}{0.48\textwidth}

    \centering
    \includegraphics[width=0.85\textwidth]{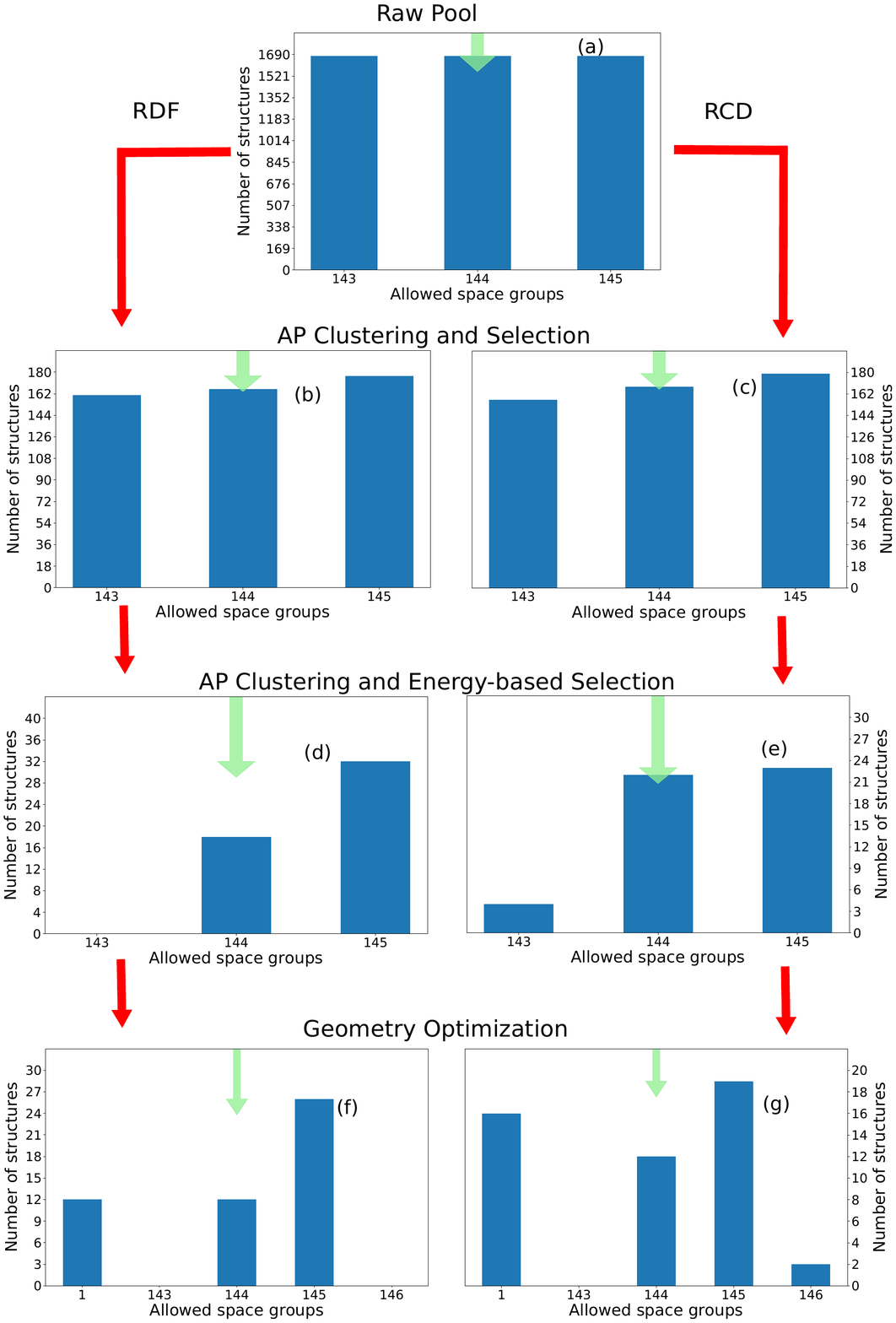}
    \caption{Space group distribution histograms obtained at each step of the Robust workflow for glycine with $Z=3$. The green arrow points to the space group of the experimental structure, $P3_1$ (144).}
    \label{fig:gly_3mpc_spg}

\end{minipage} \hfill
\begin{minipage}{0.48\textwidth}

    \centering
    \includegraphics[width=0.85\textwidth]{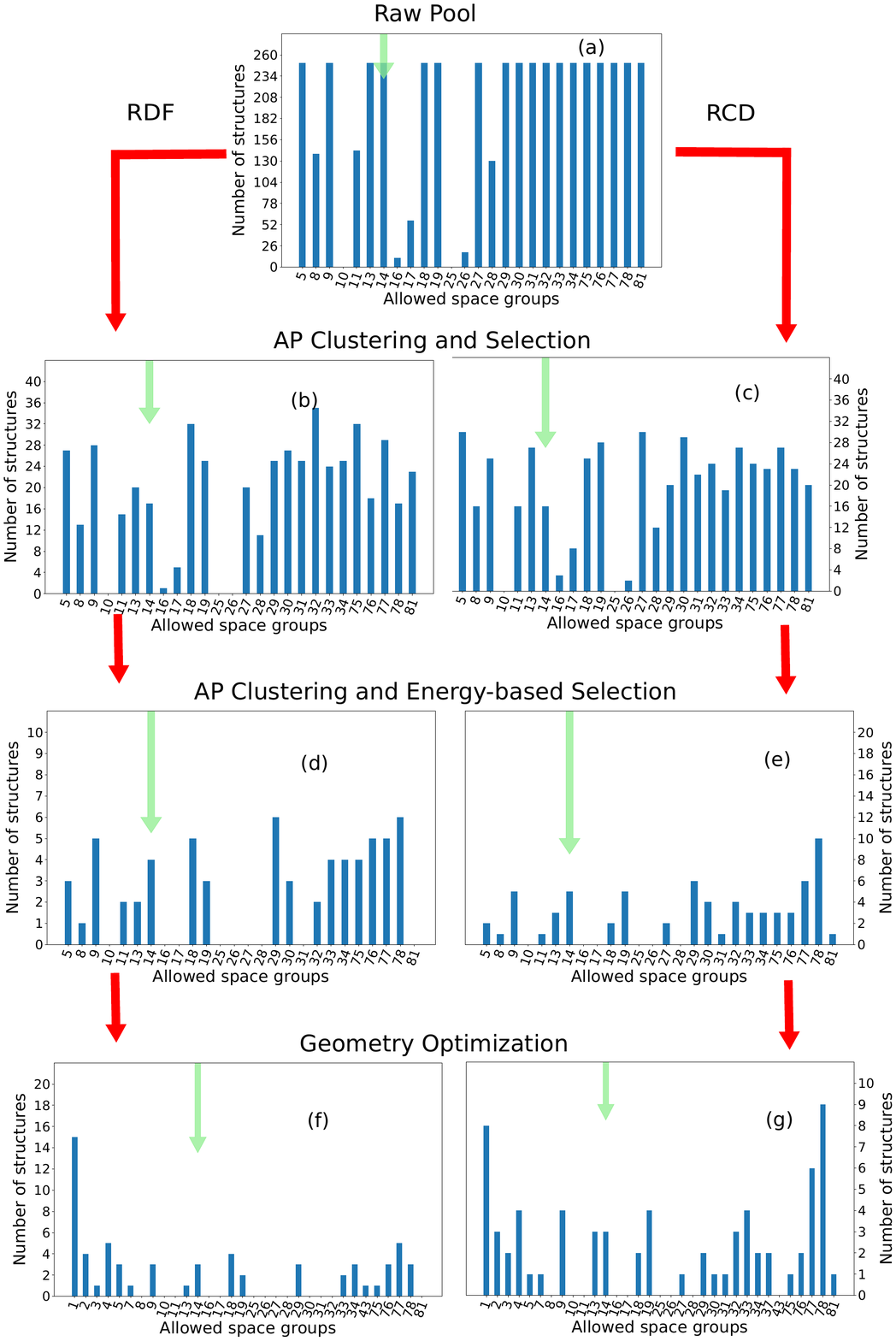}
    \caption{Space group distribution histograms obtained at each step of the Robust workflow for glycine with $Z=4$. The green arrow points to the space group of the experimental structure, $P2_1/n$ (14).}
    \label{fig:gly_4mpc_spg}

\end{minipage}
\end{figure}

\begin{figure}[h]
\begin{minipage}{0.48\textwidth}

    \centering
    \includegraphics[width=0.85\textwidth]{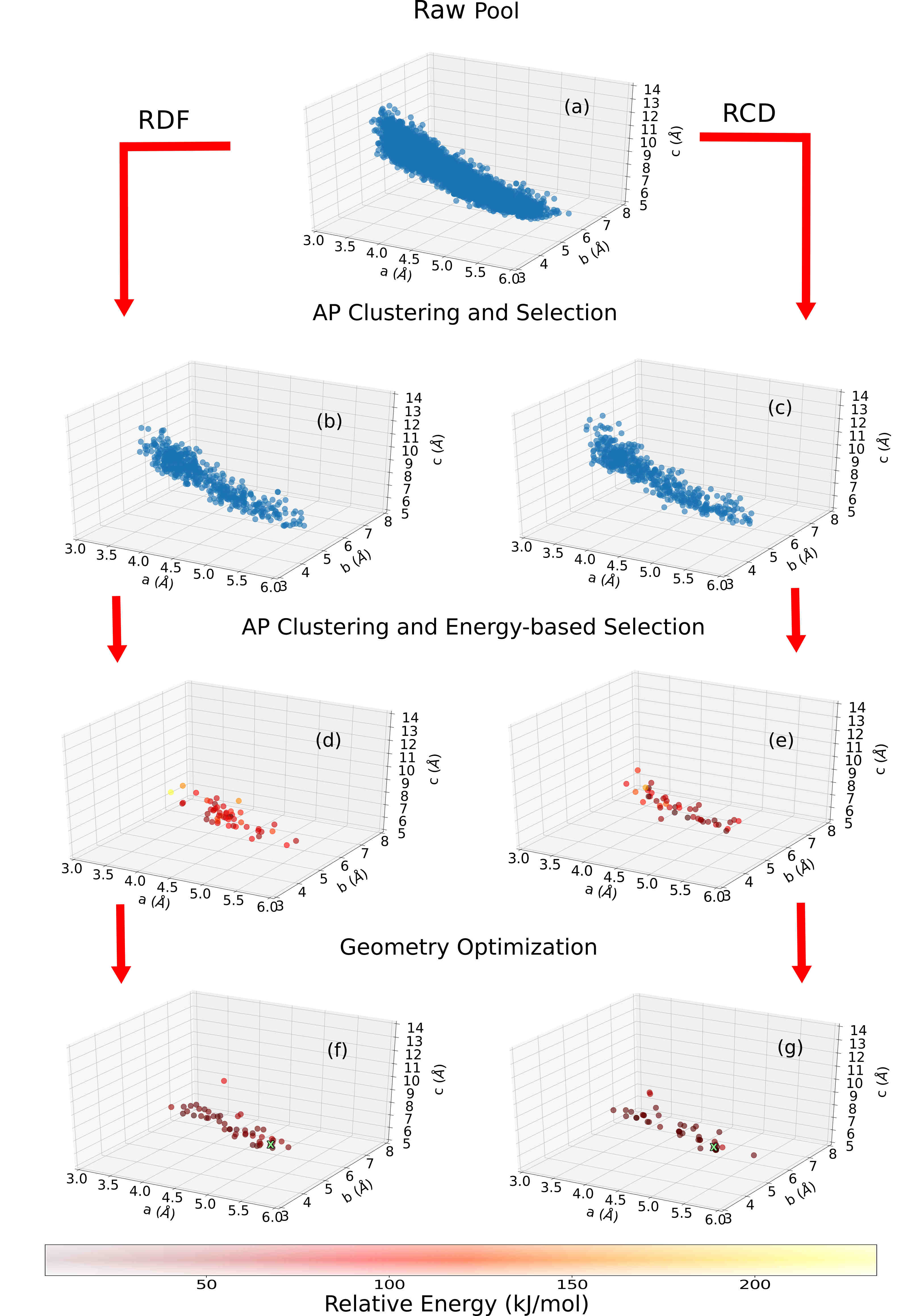}
    \caption{Lattice parameter distributions obtained at each step of the Robust workflow for glycine with $Z=2$. The green cross denotes the experimental structure.}
    \label{fig:gly_2mpc_lat}
    
\end{minipage} \hfill
\begin{minipage}{0.48\textwidth}

    \centering
    \includegraphics[width=0.85\textwidth]{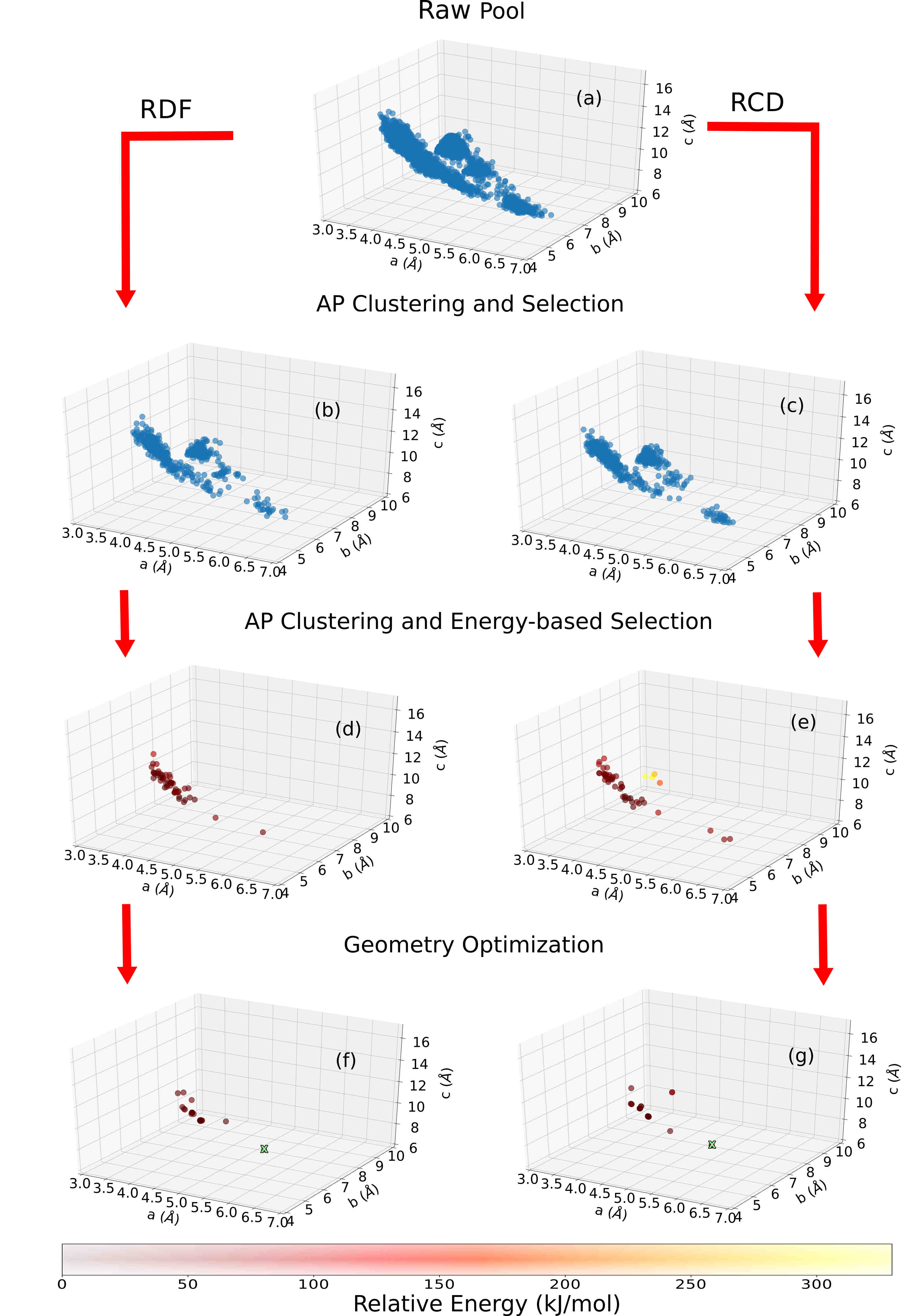}
    \caption{Lattice parameter distributions obtained at each step of the Robust workflow for glycine with $Z=3$. The green cross denotes the experimental structure.}
    \label{fig:gly_3mpc_lat}

\end{minipage}
\end{figure}
\begin{figure}[h]
    \centering
    \includegraphics[width=0.5\textwidth]{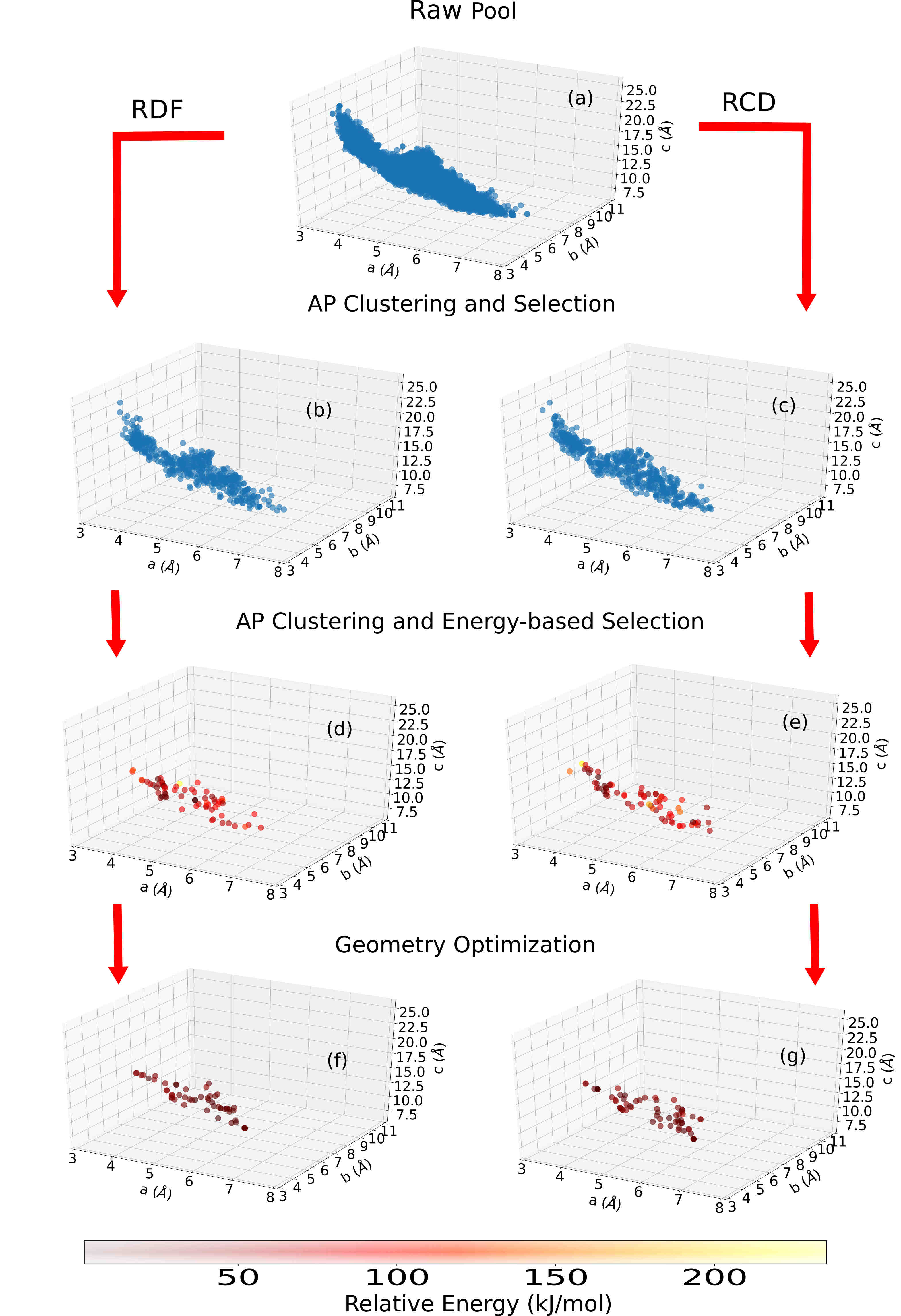}
    \caption{Lattice parameter distributions obtained at each step of the Robust workflow for glycine with $Z=4$.}
    \label{fig:gly_4mpc_lat}
\end{figure}

Glycine is the simplest proteinogenic amino acid. It is achiral and forms a zwitterion in the solid state. Therefore, the zwitterionic form of glycine was used in the single molecule relaxation step. Under ambient conditions, glycine has three common polymorphs: a) $\alpha$-glycine with $Z=4$ and space group $P2_1/n$ (14), b) $\beta$-glycine with $Z=2$ and space group $P2_1$ (4), and c) $\gamma$-glycine with $Z=3$ and space group $P3_1$ (144)/ $P3_2$ (145) \cite{boldyreva2003structural}. The structures belonging to the two space groups are enantiomorphic forms of the chiral $\gamma$-glycine crystal. Experimentally, it has been found that the relative thermodynamic stability of the polymorphs at room temperature follows $\gamma > \alpha > \beta $ with Gibbs free energy difference ($\Delta G$) of $0.16$ kJ/mol between $\gamma$-glycine and $\alpha$-glycine \cite{boldyreva2003polymorphism}. At temperatures higher than $440$ K, $\alpha$-glycine becomes more stable than $\gamma$-glycine. The crystal structure and relative stabilities of the glycine polymorphs have been studied extensively, using different computational methods \cite{chisholm2005ab, day2005beyond, marom2013many, zhu2012constrained, lund2015crystal, sabatini2012structural, rodriguez2019structural}. 

Glycine is known for its ability to form strong intermolecular hydrogen bonds, owing to which it crystallizes in a relatively dense molecular solid. Figures \ref{fig:gly_2mpc_vol}, \ref{fig:gly_3mpc_vol}, and \ref{fig:gly_4mpc_vol} show the volume histograms for $Z=2$, $Z=3$, and $Z=4$, respectively, at each step of the Robust workflow. The experimental volume and the predicted volume are indicated by solid green and dashed orange lines, respectively. The predicted volume per molecule for glycine is $79 \AA^3$, which is close to the experimental value for all polymorphs. About 5000 structures with mean unit cell volume and standard deviation of $(159, 238, 318)$ $\AA^3$ and $(12, 18, 24)$ $\AA^3$, respectively, were generated for the $Z = (2,3,4)$ polymorphs. The generated raw pool for $Z=3$ is approximately Gaussian centered on the predicted volume. Whereas for $Z=2$ and $Z=4$, the mean of the distribution is larger than the predicted volume. This is because the $Z=3$ is much easier to generate as two out of the three space groups that are allowed have screw axes. The new settings for hydrogen-bonded systems helped generate much dense structures that are close to the predicted volume. Panels (d) and (e) in Figures \ref{fig:gly_2mpc_vol}, \ref{fig:gly_3mpc_vol}, and \ref{fig:gly_4mpc_vol} show that energy-based selection and the final relaxation favor structures near the experimental volume. 

Figures \ref{fig:gly_2mpc_spg}, \ref{fig:gly_3mpc_spg}, and \ref{fig:gly_4mpc_spg} show the space group distribution for each step of the Robust workflow for glycine with $Z=2$, $Z=3$, and $Z=4$, respectively. The raw pools for all cases show almost uniform space group  distribution. For $Z=4$, space groups $P2/m$ (10) and $Pmm2$ (25) are missing because they contain mirror planes that are hard to generate \cite{pidcock2003database}. There are a significant number of structures in the experimental space group in the raw pool and subsequently selected pools for all the cases. Relaxation of the final pool may break existing symmetries or create new ones as there are no constraints imposed. This resulted in additional space groups with a  different $Z$ or $Z'$. For example, space group $Cmc2_1$ (36) and space group $P1$ were created after geometry optimization for glycine with $Z=2$ as shown in panels (f) and (g) of Figure \ref{fig:gly_2mpc_spg}.

Figures \ref{fig:gly_2mpc_lat}, \ref{fig:gly_3mpc_lat}, and \ref{fig:gly_4mpc_lat} show the lattice parameter distributions obtained at each stage of the Robust workflow for glycine with $Z$ = $2$, $3$, and $4$, respectively. For the energy-based selection and final relaxation steps, the color scale corresponds to relative energies with respect to the lowest energy structure in the final relaxed pool.  For $Z=2$ and $Z=4$, the lattice parameter space is well-sampled and diverse regions are obtained upon down-selection. For $Z = 3$, the generated structures are concentrated in  distinct regions of the lattice parameter space because there are only three compatible space groups, all of which are in the hexagonal crystal family. For the $Z=2$ case, the experimental structure of $\beta$-glycine  was found in the final relaxed pool for both RDF and RCD runs. Similarly, $\gamma$-glycine was found in both runs with $Z=3$. Although $\alpha$-glycine was not found for both the runs, a few structures with similar packing motif, same space group, or similar lattice parameters of experimental structure were found in the final pool. We note that the goal of the Robust workflow is not necessarily to find the experimental structure but to produce an initial population for a structure search algorithm, such as a genetic algorithm. 

\section{Conclusion}
\label{section:Conclusion}

In summary, we have presented a new version of the molecular crystal random structure generator, Genarris, with several new features and demonstrated its application to benzene and glycine. The new MPI parallelization scheme has made Genarris 2.0 significantly faster than the previous version, more portable, and able to scale better on high performance computing architectures. The new machine learning method for volume estimation has been demonstrated to reliably predict the volumes of the polymorphs of benzene and glycine. The somewhat larger deviation from the experimental volume for the high-pressure polymorph of benzene was expected, considering that the model was trained on crystal structures obtained at ambient pressure.

For all polymorphs of benzene and glycine, the new structure generation function has successfully generated structures in the target volume range with approximately uniform space group distributions and has adequately sampled the possible range of lattice parameters. The new capability to generate structures with molecules occupying special Wyckoff positions has proven to be instrumental for benzene. The updated structure check settings for strong hydrogen bonds have been particularly useful for glycine. Thus, Genarris 2.0 is expected to deliver a significantly better performance than the previous version for symmetric molecules and for molecules capable of forming strong hydrogen bonds. 

A new “Robust” workflow has been implemented for clustering and down-selection of the raw pool of random structures to form a small curated population of low-energy structures with diverse crystal packing motifs. Although this workflow is intended for producing an initial population for other structure search algorithms (such as genetic algorithms), not as a structure prediction method, the experimental structures of both polymorphs of benzene and of the beta and gamma forms of glycine were found in the final relaxed pools. For alpha glycine, the final pools contained several structures in the same space group, with similar lattice parameters, or with similar packing motifs, which should be sufficient for a genetic algorithm to generate the experimental structure.

Genarris 2.0 offers the user full flexibility to design and easily implement new workflows by sequentially executing a user-defined list of procedures. For example, to generate datasets for training machine learning models, the user may wish to perform energy evaluations for a larger number of structures from the raw pool. To perform crystal structure prediction, the user may wish to fully relax a larger number of structures. Thus, Genarris 2.0 is a useful random structure generator for homomolecular crystals of semi-rigid molecules with no rotatable bonds, which can be applied to generate initial populations for structure search algorithms or to generate datasets for machine learning or as a standalone crystal structure prediction method.

\section*{Acknowledgements}

Work at CMU was funded by the National Science Foundation (NSF) Division of Materials Research through grant DMR-1554428. An award of computer time was provided by the Innovative and Novel Computational Impact on Theory and Experiment (INCITE) program. This research used resources of the Argonne Leadership Computing Facility, which is a DOE Office of Science User Facility supported under Contract DE-AC02-06CH11357 and of the National Energy Research Scientific Computing Center (NERSC), a U.S. Department of Energy Office of Science User Facility operated under Contract No. DE-AC02-05CH11231. Dr. William Paul Huhn from ALCF is thanked for his help with compiling and importing FHI-aims as a Python library.



\nocite{*}
\bibliographystyle{model1a-num-names.bst}
\bibliography{references}







\end{document}